\documentclass[lettersize,journal]{IEEEtran}
\usepackage{amsmath,amsfonts}
\usepackage{algorithmic}
\usepackage{array}
\usepackage{textcomp}
\usepackage{stfloats}
\usepackage{url}
\usepackage{verbatim}
\usepackage{graphicx}
\usepackage{cite}
\usepackage{xcolor}
\usepackage{ntheorem}
\usepackage{caption}
\usepackage{makecell}
\usepackage{booktabs}
\usepackage{enumitem}

\usepackage{multirow}
\usepackage{pdfpages} 
\usepackage{makecell}
\usepackage{subcaption}
\usepackage[bookmarks=true,breaklinks=true,letterpaper=true,colorlinks,linkcolor=blue,citecolor=magenta,urlcolor=blue]{hyperref}
\usepackage[ruled,noline,noend]{algorithm2e}

\usepackage{lipsum}

\begin{document}

\title{A Survey of Real-time Scheduling on Accelerator-based Heterogeneous Architecture for Time Critical Applications}


\author{An Zou$^{1}$, Yuankai Xu$^{1}$, Yinchen Ni$^{1}$, Jintao Chen$^{1}$, Yehan Ma$^{1}$, Jing Li$^{2}$, \\ Christopher Gill$^{3}$, Xuan Zhang$^{4}$, Yier Jin$^{5}$ \\ $^{1}$Shanghai Jiao Tong University, $^{2}$New Jersey Institute of Technology, \\ $^{3}$Washington University in St. Louis, $^{4}$Northeastern University, $^{5}$University of Science and Technology of China 
}

\maketitle

\begin{abstract}
Accelerator-based heterogeneous architectures—such as CPU-GPU, CPU-TPU, and CPU-FPGA systems—are widely adopted to support the popular artificial intelligence (AI) algorithms that demand intensive computation. When deployed in real-time applications, such as robotics and autonomous vehicles, these architectures must meet stringent timing constraints. To summarize these achievements, this article presents a comprehensive survey of real-time scheduling techniques for accelerator-based heterogeneous platforms. It highlights key advancements from the past ten years, showcasing how proposed solutions have evolved to address the distinct challenges and requirements of these systems.

This survey begins with an overview of the hardware characteristics and common task execution models used in accelerator-based heterogeneous systems. It then categorizes the reviewed works based on soft and hard deadline constraints. For soft real-time approaches, we cover real-time scheduling methods supported by hardware vendors and strategies focusing on timing-critical scheduling, energy efficiency, and thermal-aware scheduling. For hard real-time approaches, we first examine support from processor vendors. We then discuss scheduling techniques that guarantee hard deadlines (with strict response time analysis). After reviewing general soft and hard real-time scheduling methods, we explore application- or scenario-driven real-time scheduling techniques for accelerator-enabled heterogeneous computing platforms. Finally, the article concludes with a discussion of open issues and challenges within this research area.
\end{abstract}


\section{Introduction}
\label{sec:intro}
The computing systems, no matter for embedded or edge applications, are heading towards heterogeneity to support the intensive computation in emerging artificial intelligence (AI) tasks \cite{anderson2019extreme}. These artificial intelligence tasks, in many real-world scenarios such as autonomous driving \cite{redmon2017yolo9000} and robotics \cite{michel2007gpu}, are facing strict timing constraints. 
The heterogeneous computing platforms, such as NVIDIA and AMD GPUs \cite{choquette2020nvidia}, Xilinx UltraScale \cite{leibson2013xilinx}, and TI Keystone II \cite{schwaller2017investigating} SoCs blend CPU cores and accelerator parallel processing elements (PEs), such as GPU Streaming Multiprocessors, in one architecture.

\begin{figure*}
\centering
\includegraphics[width=0.65\textwidth]{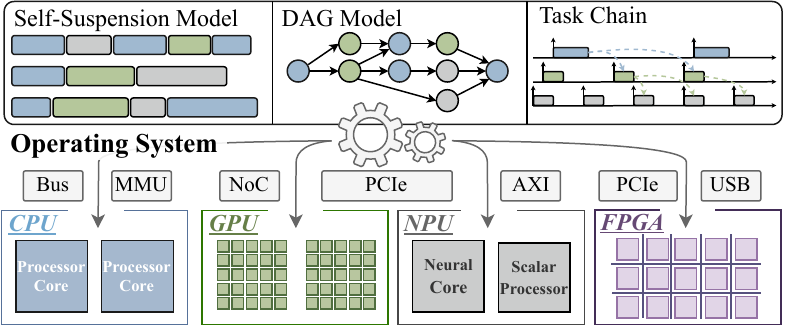}
\caption{Real-time scheduling of parallel tasks on the heterogeneous architecture.}
\label{fig:intro}
\end{figure*}

The tasks on accelerator-based heterogeneous computing architecture exhibit a segmented structure, as illustrated in Fig. \ref{fig:intro}. To optimize performance and energy efficiency, serial computation segments within a task are typically allocated to CPU cores, referred to as "CPU workloads or CPU segments." In contrast, data-parallel segments, labeled "accelerator workloads or accelerator segments," are well-suited for offloading to the accelerator processing elements (PEs). This segmented execution pattern aggravates the dependencies and competition between parallel tasks \cite{brightwell2018resource}. Additionally, the complex task execution patterns make it challenging to simultaneously meet timing constraints while achieving high resource utilization rates \cite{chen2019many,huang2016self}. Significant reductions in schedulability are often observed when multiple types of accelerators coexist, with an increasing number of CPU cores and accelerator PEs \cite{patel2018analytical}, as well as a growing number of corresponding computation segments \cite{saha2019stgm}.

The research on real-time scheduling for heterogeneous architectures focuses on accelerator-based systems, where the accelerators can be GPUs, TPUs, or FPGAs. This body of survey is driven by three main objectives. 

The first objective addresses soft real-time tasks, which typically involves integrating real-time schedulers within both the operating system and the hardware architecture of heterogeneous cores. These scheduler designs often leverage heuristic approaches \cite{wang2021balancing} that capitalize on emerging hardware features. Such approaches are particularly suited for real-time applications, where the emphasis is on maximizing schedulability and tolerating occasional extreme cases. Since these solutions target soft real-time applications, detailed response time analysis (i.e., determining time boundaries) is not always required for many scheduler designs.

The second objective targets hard real-time tasks that require guaranteed end-to-end response times to ensure schedulability. In parallel task execution on heterogeneous computing platforms, increased inter-task dependencies and resource contention often introduce significant pessimism in response time analysis. Many existing works rely on traditional scheduling techniques, such as fixed-priority or earliest-deadline-first (EDF) scheduling, and aim to derive tight or even exact response time bounds.

In addition to research aimed at enhancing soft and hard real-time performance for general-purpose applications and architectures, there is also a body of work dedicated to improving real-time performance for specific applications.

\begin{figure*}
\centering
\includegraphics[width=0.8\textwidth]{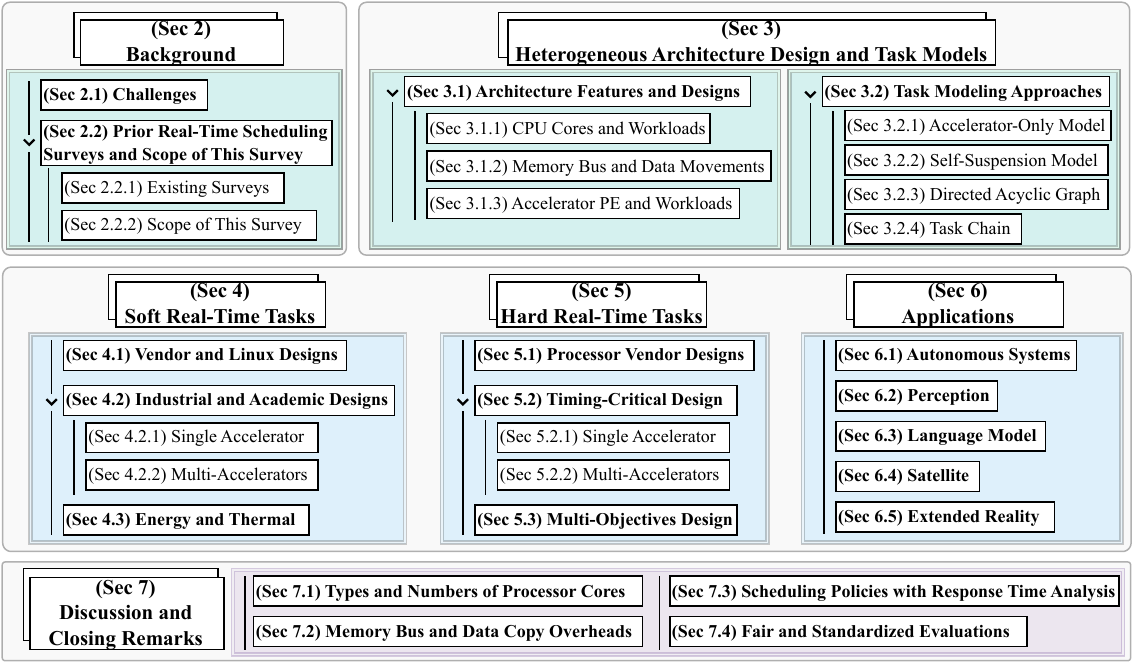}
\caption{Overview of this survey.}
\label{fig:overview}
\end{figure*}

Therefore, in this paper, we survey the research on real-time scheduling for accelerator-based heterogeneous architectures, structured around three key areas: scheduling for soft real-time tasks, hard real-time tasks, and application-driven real-time scheduling, as overviewed in Fig. \ref{fig:overview}. Section \ref{sec:background} introduces the challenges, previous surveys, and scopes associated with real-time scheduling on heterogeneous architectures. Section \ref{sec:model} starts this survey with the architectural designs and commonly used models. Section \ref{sec:softdeadline} reviews the real-time scheduling approaches for soft real-time tasks, while Section \ref{sec:harddeadline} surveys the work on hard real-time tasks. Section \ref{sec:application} explores application- and scenario-driven real-time scheduling approaches. Finally, Section \ref{sec:remark} discusses potential future directions for time-critical computing and real-time scheduling on the accelerator-enabled heterogeneous architectures.
\section{Background}
\label{sec:background}
\subsection{ Challenges}

Accelerator-based heterogeneous systems, which combine CPU cores and accelerator processing elements (PEs), are becoming increasingly prevalent, from embedded computing platforms (e.g., NVIDIA Jetson Series) to high-performance computing environments (e.g., Oak Ridge's Titan supercomputer). These systems can offer superior performance compared to conventional homogeneous systems by the benefits of parallel computing from accelerator PEs.

In such systems, CPU cores serve as the central controllers, while accelerator PEs are regarded as auxiliary devices designed to accelerate parallel arithmetic operations. Typically, for an application running on a heterogeneous system, the CPU handles I/O and serial computation, while parallel tasks are offloaded to the accelerator PEs.

The key challenges of time-critical computing and real-time scheduling on the heterogeneous architecture are to simultaneously deal with the non-unified and inflexible access patterns to diverse processors; deal with the internal dependence between segments in one task and the external parallelism between each task; deal with the competition on the limited hardware resources and the under-utilization due to the inherent task affinity. These challenges are summarized in the following three points:

\begin{itemize}
    \item \textbf{Non-unified and Inflexible Access Patterns:} Unlike the common features of preemption and core-level affinity in CPUs, accelerators exhibit non-unified and inflexible access management mechanisms for preemption and partitioning. Since accelerators are primarily designed for maximum throughput per chip area, many of them are hard to support preemption and partition, except for a few that support limited forms of partitioning or preemption. This non-unified access pattern significantly complicates scheduling research, while the inflexible nature of these patterns severely hinders scheduling performance.
    \item \textbf{Parallel Tasks with Serial Dependencies:} The task set consists of multiple parallel tasks, each containing a series of dependent segments that need to be scheduled in a specific order. The challenge arises from the need to account for both the parallelism of independent tasks and the interdependencies between task segments, which must be executed sequentially or in a specific sequence.
    \item \textbf{Resource Contention with Under-utilization:} While multiple tasks compete for accessing to limited hardware resources, such as CPU cores and processing elements (PEs), the inherent task affinity (e.g., CPU segments running on CPU cores, accelerator segments running on accelerator PEs) and the dependencies (e.g. accelerator segment must be executed after its previous CPU segment) between task segments often prevent full utilization of available hardware. This results in resource contention and suboptimal resource utilization, even though there is available capacity.
\end{itemize}

To address the challenges outlined above, this survey begins by examining the various heterogeneous architecture designs and task models that form the foundation for effective time-critical computing and scheduling. Then the real-time scheduling of soft real-time tasks, hard real-time tasks, and the application-driven scheduling are presented.

\subsection{Prior Real-Time Scheduling Surveys and Scope of This Survey}
\subsubsection{Existing Surveys} 
Previous works have provided well-crafted surveys that are relevant to, yet distinct from, our study.
(1) Several prior surveys have investigated scheduler designs for multicore homogeneous processors, covering aspects such as safety \cite{cerrolaza2020multi}, schedulability \cite{davis2011survey, maiza2019survey}, performance \cite{singh2017survey, mittal2016survey}, and energy efficiency \cite{bambagini2016energy, sheikh2018energy}.
(2) In addition to general-purpose homogeneous processors, surveys have also addressed real-time scheduling on limited preemptive processors \cite{buttazzo2012limited} and implementations of real-time scheduling in the Linux kernel \cite{reghenzani2019real}.
(3) For artificial intelligence applications, earlier works have summarized scheduling and load-balancing strategies \cite{YeDeep2023, liang2024communication}, as well as neural network training techniques \cite{duan2024efficient, liang2024resource}, primarily in the context of cloud-based (virtualized) GPU servers where timing constraints are weak or even negligible.

\subsubsection{Scope of This Survey and Relevant Studies Not Included in This Survey}
\textit{\textbf{This work focuses on the scheduling of real-time tasks on accelerator-enabled heterogeneous computing platforms. A computation task is considered a real-time task (for timing-critical applications) if it has hard or soft deadlines, which serve as the primary metric for inclusion in this survey.}} Due to space and scope constraints, we do not cover latency- or quality-of-service (QoS)-driven scheduling approaches in this paper.

While our survey is not limited to single-machine heterogeneous architectures, most existing research on real-time scheduling for time-critical applications tends to concentrate on such systems rather than cloud-based (virtualized) accelerator servers. This focus is largely attributed to two factors: (1) real-time applications are more commonly found in embedded or mobile environments, and (2) the inherent timing unpredictability introduced by virtualization can easily compromise real-time guarantees.

\section{Heterogeneous Architecture Designs and Task Models}
\label{sec:model}

\subsection{Architecture Features and Designs}

\subsubsection{CPU Cores and Workloads}
As illustrated in Fig. \ref{fig:intro}, applications on heterogeneous computing platforms typically consist of multiple segments: CPU workloads, memory copies between CPU cores and accelerator processing elements (PEs), and accelerator workloads. Due to their powerful parallel computational capabilities, accelerators are typically assigned computationally intensive tasks such as matrix operations. The CPU, on the other hand, is responsible for executing serial instructions, such as interfacing with I/O devices (e.g., sensors and actuators), as well as initiating memory transfers and accelerator tasks.

Once the CPU dispatches memory copies and accelerator segments into a FIFO buffer, it can immediately proceed to execute subsequent instructions—unless explicit synchronization mechanisms are used to wait for the completion of these tasks. Consequently, CPU segments are usually modeled as serial instructions executed by a single thread. Meanwhile, mainstream CPU architectures, such as x86, ARM, and RISC-V, commonly support interrupts and preemption, enabling responsive and flexible task scheduling on the CPU side.

\subsubsection{Memory Bus and Data Movements}
Data movements copying between the CPU cores and accelerator PEs include two stages. In the first stage which is also called global memory copy (a terminology from NVIDIA), data is copied between the CPU memory and the accelerator memory through the memory bus. The advanced extensible interface (AXI), network on chip (NoC), and single peripheral component interconnect express (PCIe) are the most common bus in the embedded and desktop/server heterogeneous computing architecture. The all these three can offer packet-based and full-duplex communication between any two endpoints. The number of global memory copies that can happen simultaneously is determined by the number of copy engines specified by the bus of the heterogeneous architectures. For example, GeForce GTX TITAN Black GPU and Jetson TX2 SoC have 1 copy engine, while 1080 TI, TITAN X GPU, and NVIDIA Xavier SoC have 2 copy engines. The global memory copy through these three buses in the heterogenous architectures is generally non-preemptive once it starts.
The accelerators could provide two types of global memory movement \cite{amert2017gpu, otterness2017evaluation}: explicit memory copy and implicit memory copy (also called zero-copy memory). Explicit memory copy uses traditional memory, where data must be explicitly copied from CPU to PE portions of DRAM. Unified memory is developed from zero-copy memory where the CPU cores and the accelerator PEs can access the same memory area by using the same memory addresses between the CPU cores and accelerator PEs. The real-time scheduling approaches designed for explicit memory copies can be directly applied to implicit memory copies by setting the explicit copy length to zero.

The second stage is the memory access from the accelerator PE’s execution units to the PE cache (also called buffers) or memory. Most accelerators adopt a hierarchical memory architecture. These memory accesses happen simultaneously with the instruction execution on PEs. Compared to the global memory copy, the second stage of memory operation can be measured and modeled as part of the PE execution model. Although run-time memory factors, such as the state of the row buffers in the first stage and contention on memory or cache in the second stage, would impact memory copy time and worst-case execution time WCET \cite{chen2015execution,yandrofski2022making,bakita2024demystifying}, the end-to-end scheduling may choose to simplify the memory model with static factors, given the consideration of real-time scheduling complexity \cite{betts2013estimating}.

\subsubsection{Accelerator PE and Workloads}

\begin{tiny}
\begin{table*}[]
\renewcommand{\arraystretch}{1.5}
\caption{Summary of accelerator architecture designs to support real-time computing (Here, we summarize the works mainly on the system-level designs to improve the flexibility of accelerator cores. The approaches that work on scheduling and system co-designs will be introduced in later scheduling sections).}
\begin{tabular}{|l|l|l|l|l|l|}
\hline
\textbf{}                                                                                                   & \textbf{\begin{tabular}[c]{@{}l@{}}Approach \\ Name\end{tabular}} & \textbf{\begin{tabular}[c]{@{}l@{}}Preemption/\\ Partitioning\end{tabular}} & \textbf{\begin{tabular}[c]{@{}l@{}}Software/ Hardware \\ Approaches \end{tabular}} & \textbf{\begin{tabular}[c]{@{}l@{}}Other \\ Approaches\end{tabular}} & \multicolumn{1}{l|}{\textbf{\begin{tabular}[c]{@{}c@{}}Features\end{tabular}}} \\ \hline
\multirow{7}{*}{\textbf{\begin{tabular}[c]{@{}l@{}}GPU\end{tabular}}} 

& Elliott \cite{elliott2013gpusync} & Preemption & \begin{tabular}[c]{@{}l@{}} Software by  \end{tabular} & N/A & Thread block level preemption \\ \cline{2-6} 

& Chimera \cite{park2015chimera} & Preemption & \begin{tabular}[c]{@{}l@{}}Hardware on simulator \end{tabular} & N/A & Thread block level preemption \\ \cline{2-6} 

& Wang et, al.\cite{wang2024unleashing} & Preemption & \begin{tabular}[c]{@{}l@{}} Software by device driver \end{tabular} & N/A & Thread block level preemption \\ \cline{2-6} 
                                                                                                            
& Basaran et, al. \cite{basaran2012supporting} & Preemption & Software & N/A & Kernel level preemption \\ \cline{2-6} 
                                                                                                            
& Tanasic et, al.\cite{tanasic2014enabling} & Preemption & \begin{tabular}[c]{@{}l@{}}Hardware on simulator \end{tabular} & N/A & Kernel level preemption \\ \cline{2-6} 

& GCAPS\cite{wang_et_al:LIPIcs.ECRTS.2024.14} & Preemption & Software by input/output control& N/A & Kernel level preemption\\ \cline{2-6}  

& NVIDIA MIG \cite{choquette2021nvidia} & Partitioning & Hardware & N/A & Official Support on advanced GPU \\ \cline{2-6}

& Bakita et, al. \cite{bakita2023hardware} & Partitioning & Hardware & N/A & Support on NVIDIA GPU since 2013 \\ \cline{2-6}
\hline
\multirow{4}{*}{\textbf{\begin{tabular}[c]{@{}l@{}}TPU \\ (Systolic \\ Array) \end{tabular}}}     
                    & SEPT\cite{han2023spet} & Preemption & Hardware &  SRAM &  Network layer level preemption \\ \cline{2-6} 
                    & PREMA\cite{choi2020prema} & Preemption & Hardware & N/A & Kernel level preemption  \\ \cline{2-6} 
                    & Dataflow-Mirroring\cite{choi2023enabling} & Partitioning & Hardware & N/A & Multi Tasks Parallelization \\ \cline{2-6} 
                    & Reshadi\cite{reshadi2023dynamic} & Partitioning & Hardware &  Buffers & Partition to multi-tenants \\ \hline
\multirow{2}{*}{\textbf{\begin{tabular}[c]{@{}l@{}}ASIC \\ (FPGA)\end{tabular}}}     & FRED\cite{biondi2016fred} & Partitioning & Software by device driver & With RTA & Scheduling both HW and SW tasks \\ \cline{2-6} 
                     & Cordone et al.\cite{cordone2009partition} & Partitioning & Software by device driver & N / A & Solving partitioning scheme by LP \\ \cline{2-6}
                    & Hoornaert et al.\cite{hoornaert2021memory} & Memory control & kernal and hardware codesign & N / A & Fine-grained memory access control \\ \cline{2-6} 
                    & Roozkhosh et al.\cite{roozkhosh2020potential} & Cache partitioning  & Fine-grained memory transaction & N / A & Cache partitioning \\ \cline{2-6} \hline
\end{tabular}
\label{tab:system}
\end{table*}
\end{tiny}

Heterogeneous accelerators leverage parallel processing elements (PEs) to accelerate computations that can be parallelized. These PEs can operate independently or collaboratively as a cluster. For example, in FPGA-based heterogeneous architectures, IP cores (i.e., PEs) can typically function independently. In contrast, in GPU-based architectures, streaming multiprocessors (SMs) generally operate as clusters by default. 

To achieve high performance under strict power and area constraints, most accelerators forgo the auxiliary circuitry required to support preemption. Although numerous studies have demonstrated that system-wide schedulability can benefit from preemption support, such capabilities are rarely implemented in PE hardware. Instead, most preemption mechanisms for PEs are realized through software techniques. In the context of GPUs, Park~\cite{park2015chimera}, Basaran~\cite{basaran2012supporting}, Tanasic~\cite{tanasic2014enabling}, and Zhou~\cite{zhou2015gpes} proposed architectural extensions using hardware/software co-designs to enable preemption, evaluating their approaches on GPU simulators. The Effisha framework~\cite{chen2017effisha} introduced a purely software-based solution for supporting kernel preemption at the granularity of arbitrary thread block boundaries, without requiring hardware modifications. Similarly, for FPGAs, Rodriguez-Canal~\cite{rodriguez2022programming} introduced programming abstractions for preemptive scheduling through dynamic partial reconfiguration, enabling finer control over task execution without redesigning the hardware.

Alongside temporal preemption, spatial partitioning enhances both access flexibility and schedulability of clustered processing elements (PEs). In recent years, both researchers and processor vendors have increasingly supported spatial partitioning to enable concurrent applications. For instance, NVIDIA introduced Multi-Process Service (MPS)\cite{gandham2021improving} and Multi-Instance GPU (MIG)\cite{choquette2020nvidia}, which allow multiple tasks to run concurrently by assigning specific numbers of PEs to each task. Similarly, AMD released open-source software support for hardware partitioning, which is expected to accelerate progress and contribute to the long-term viability of real-time GPU research~\cite{otterness2021exploring, otterness2020amd}. In addition, researchers have proposed architectural support for spatial partitioning in systolic arrays~\cite{choi2023enabling}, which serve as accelerators for general matrix multiplication and convolution operations. These architectural advancements, aimed at improving accelerator flexibility and supporting real-time performance, are summarized in Table~\ref{tab:system}.

Therefore, the features of accelerator-based heterogenous architectures can be summarized to comprise the following key features.

1. CPU cores operating \textbf{preemptively}.

2. Memory copies function in a \textbf{non-preemptive} manner.

3. The accelerator and its PEs (computing units in the accelerator) can operate using one of these approaches: \textbf{non-preemptive and non-partitioning}, \textbf{preemptive}, or \textbf{partitioning}.

In the following sections, we uniformly refer to the cores in heterogeneous processors (such as GPU streaming multiprocessors and FPGA IP cores) as processing elements (PEs). Accordingly, the jobs executed on the CPU, memory bus, and PEs are referred to as CPU, memory, and accelerator segments, respectively.

\subsection{Task Modeling Approaches}

\begin{figure*}[t]
\centering
\begin{subfigure}{0.32\textwidth}
    \includegraphics[clip, width=1\linewidth]{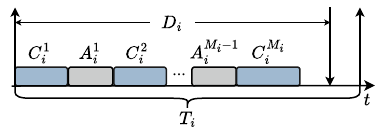}
    \caption{SSSM.}
    \label{fig:SSSM} 
\end{subfigure}
\begin{subfigure}{0.32\textwidth}
    \includegraphics[clip, width=1\linewidth]{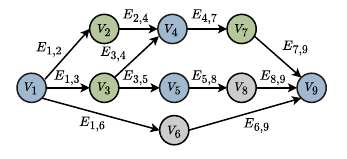}
    \caption{DAG.}
    \label{fig:DAG} 
\end{subfigure}
\begin{subfigure}{0.32\textwidth}
    \includegraphics[clip, width=1\linewidth]{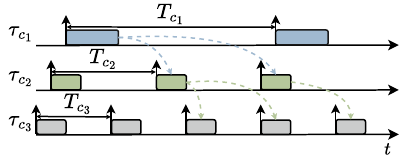}
    \caption{Task Chain.}
    \label{fig:chain} 
\end{subfigure}
\caption{Different task modeling approaches.}
\label{fig:modeling}
\end{figure*}

The parallel computing tasks running on the heterogeneous computing architecture is noted as,  $\tau_i$, $i\in \{1,2,3,..., N\}$, with a period of $T_i$ and a deadline $D_i$ for the task $\tau_i$. Researchers leverage different models to capture the internal dependency and execution pattern of the diverse tasksets, and we categorize them as follows.


\subsubsection{\textbf{Accelerator-Only Model}}
The simplest model for accelerator-based heterogeneous architectures is the accelerator-only model. Instead of incorporating both CPU and accelerator workloads, some studies simplify the task model by considering only the accelerator workloads for accelerator-intensive tasks \cite{zhou2018s, pang2023efficient}. This approach is suitable when there are ample CPU cores available, and the execution time of CPU segments is negligible compared to that of the accelerator segments.

\subsubsection{\textbf{Self-Suspension Segmented Model (SSSM)}}
One of the classic task models on heterogeneous architecture is
the segmented model~\cite{chen2014fixed,gunzel2022edf} as shown in Fig.~\ref{fig:SSSM},
which can be expressed as a 3-tuple, 
\begin{equation}
\small{
\tau_i = \big( (C_{i}^{1}, A_{i}^{1}, C_{i}^{2}, ..., A_{i}^{M_{i}-1}, C_{i}^{M_{i}}), D_{i}, T_{i} \big).}
\label{eq:self-suspension model}
\end{equation}
In this model, a task $\tau_i$ consists of ${M_i}$ CPU segments (CSs) and ${M_i - 1}$ accelerator segments (ASs). The worst-case execution times (WCETs) of the $m$-th CPU and accelerator segments are denoted by $C_i^m$ and $A_i^m$, respectively. In the self-suspending segmented model, with the CPU treated as the host, CPU segments are typically considered as computation phases, while accelerator segments are modeled as suspension intervals. This model generally assumes that each task utilizes a single accelerator, as few studies have explored scenarios where tasks switch between multiple types of accelerators \cite{zou2021rtgpu,xu2022shape}. The self-suspending segmented model captures only the intra-task dependencies of each $\tau_i$, making it particularly suitable for modeling parallel tasks (i.e., tasks without inter-task dependencies) that exhibit explicit sequential execution on heterogeneous architectures. Reflecting real-world applications, this model can effectively represent the inference stage of multiple convolutional neural networks (CNNs), where computation and accelerator usage are interleaved in a structured manner.

\textbf{Resource included model (RIM)} \cite{christmann2023formal,hapka2022controlling} is a special case of self-suspension model. In this model, each task $\tau_i$ is regarded as the basic unit of resource allocation. Thus, each task $\tau_i$ has an execution time denoted as $e_i^{Re}$ on a different computation resource $Re$. Then the whole task is denoted as $\tau_i = (e_i^{Re},T_i,D_i)$ with the release period $T_i$ and the deadline $D_i$.

\subsubsection{\textbf{Directed Acyclic Graph (DAG)}}

The DAG model depicts each task $\tau_i$ as a graph $G_i = (V_i, E_i)$, as well as its deadline $D_i$ and period $T_i$ \cite{verucchi2023dagsurvey}. Each vertex in $V_i$ corresponds to a subtask execution time and its processor affinity, while each directed edge represents the constraints that a subtask can only be executed after the completion of preceding nodes. Conditional nodes \cite{houssam2021hpcdag} can be inserted to represent multiple alternative execution paths following this model.

Multiple tasks ($\tau_1, \tau_2, \ldots, \tau_n$) form the taskset $\tau$, which executes on a heterogeneous architecture. Unlike the self-suspending segmented model, which cannot capture dependencies between tasks, the DAG model represents the dependency between task $\tau_i$ and task $\tau_j$ using a directed edge $E_{i,j}$ from vertex $V_i$ to vertex $V_j$ as shown in Fig.~\ref{fig:DAG}. When such dependencies exist, the individual task graphs ($G_1, G_2, \ldots, G_n$) are interconnected to form a larger graph $G$ that models the entire taskset $\tau$.

The DAG model targets the intricate intra-task and inter-task dependencies, meeting the demand of the ever-growing and complicated neural network architecture \cite{cai2018path}. For instance, the inference stage of transformer-based networks---espeacially the calculation of the attention\cite{vaswani2017attention}---matches well with the DAG model. The flexible execution order inside the DAG offers a vast design space for scheduling but also imposing more stringent
demands on task schedulers to deal with the non-trivial dependencies.


\subsubsection{\textbf{Task Chain}}
Task Chain (or called processing chain, cause-effect chain)~\cite{casini2019response,tang2020response,choi2021picas} models the subtasks executing as a chain, demonstrated in Fig. \ref{fig:chain}, which is denoted as $\Gamma_c=[\tau_{c_1},\tau_{c_2},...,\tau_{c_n}]$, where:
\begin{itemize}
    \item $[\tau_{c_1},\tau_{c_2},...,\tau_{c_n}]$ 
    describes the path of data through different subtasks by a finite sequence.
    Each job of the subtask $\tau_{c_{i+1}}$ reads the data not before it was written by the job of the previous subtask $\tau_{c_{i}}$ in the chain.
    \item $\tau_{c_i}=(e_{c_i}^{Re},T_{c_i},D_{c_i})$ is modeled by the RIM model.
\end{itemize}
While both the Self-Suspending Segmented Model (SSSM) and the Task Chain model feature an explicit serial execution order, the Task Chain model places greater emphasis on data dependencies and the potential communication latency between subtasks. As a result, the Task Chain model is widely applied in Robot Operating Systems (ROS), where data communication plays a pivotal role in ensuring proper system functioning.

\par
\textbf{Model summary} The aforementioned models are distinct and cater to different real-world scenarios. The self-suspension segmented model (SSSM) captures straightforward temporal dependencies, such as those found in the inference stage of a CNN. The Directed Acyclic Graph (DAG) model represents strong temporal dependencies between subtasks, such as the computation of transformer. Lastly, the Task Chain model is widely utilized in systems with data dependencies, such as in ROS.

\section{Real-time Scheduler Designs for Soft Real-Time Tasks}
\label{sec:softdeadline}

\begin{figure*}
\centering
\includegraphics[width=0.99\textwidth]{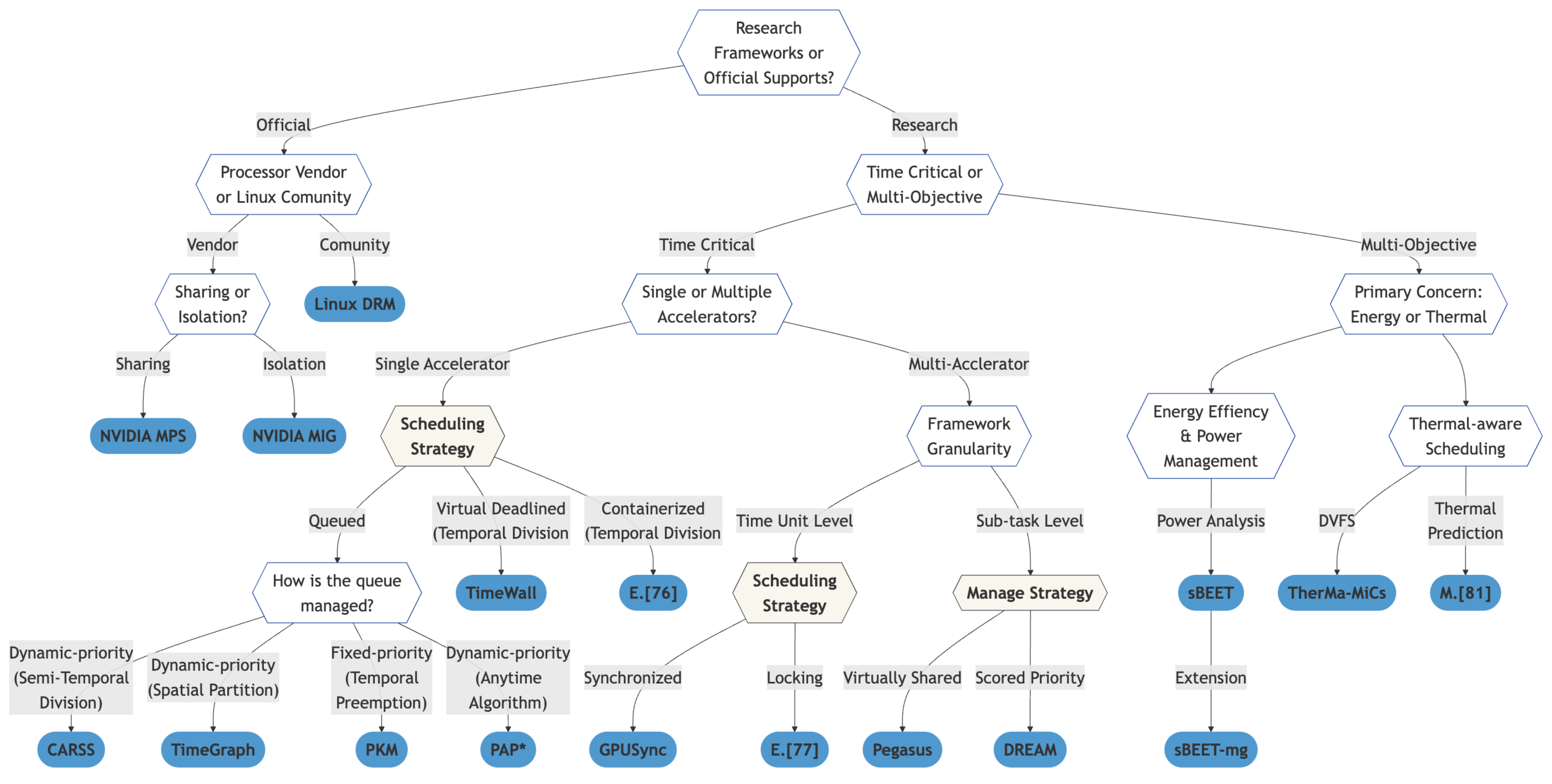}
\caption{Tree Diagram for Soft Real-Time Tasks Scheduling.}
\label{fig:srt_classify}
\end{figure*}

In this survey, we begin by presenting an overview of recent scheduling approaches for soft real-time applications, as illustrated in Fig. \ref{fig:srt_classify}. We classify these approaches as targeting soft real-time tasks, as they aim to improve system schedulability \textbf{without strict theoretical response time analysis and can tolerate occasional deadline misses.} To support the scheduling of soft real-time tasks on heterogeneous architectures, processor vendors and the Linux community provide official support to ease the deployment of real-time applications on practical systems. Following these approaches, researchers have developed scheduling strategies that target not only timing constraints but also multiple objectives, including energy efficiency, thermal management, and more.

\subsection{Designs from Processor Vendors and Linux Community}

The Linux operating system and processor vendors are supporting more and more functionalities in scheduling accelerator-based systems, the following developments are notable. The DRM (Direct Rendering Manager) GPU Scheduler \cite{drm_manage} in the Linux kernel since kernel 2.4 is designed to manage and prioritize tasks sent to the GPU for execution. It provides a fair-share scheduling mechanism that ensures multiple clients (applications) can share GPU resources efficiently. 

Apart from the general Linux scheduler, CUDA Streams and Multi-Process Service (MPS) \cite{nvidia_mps} provides basic scheduling functionality for NVIDIA GPUs, though not explicitly designed for real-time tasks. CUDA streams can prioritize tasks, while MPS allows multiple processes to share GPU resources, but neither was designed with stringent real-time requirements in mind. NVIDIA's Multi-Instance GPU (MIG) \cite{nvidia2023mig} enables the partitioning of a single GPU into multiple independent instances, each with dedicated resources, offering enhanced isolation and predictability. This provides a more suitable solution for workloads requiring resource isolation and better real-time performance compared to MPS.

Designs from processor vendors and the Linux community are general-purpose and user-friendly, but this generality also limits their effectiveness, offering only modest improvements in schedulability compared to subsequent designs proposed by industry and academic researchers.

\subsection{Designs from Researchers on Only Soft Real-Time Constraint}

In the study of time-critical schedulers for accelerator-based systems—for example CPU-GPU architectures—numerous notable works have emerged, addressing single-accelerator and multi-accelerator configurations. Importantly, the choice of target architecture (e.g., single vs. multiple accelerators) is not the sole indicator of a framework’s capability. Rather, it primarily reflects the explored design space and the specific research objectives motivating the work.

\subsubsection{Scheduling for One Accelerator Based System}

Prior approaches, such as Global Earliest Deadline First (GEDF), have been effective in providing bounded tardiness in homogeneous systems but face challenges in maintaining efficiency in heterogeneous environments. For heterogeneous systems with a single accelerator, researchers have increasingly adopted heuristically designed queuing strategies to improve real-time scheduling performance.

Temporal-based queuing is a direct and effective approach to schedule the tasks on accelerator-based heterogeneous architecture. PKM \cite{basaran2012supporting} offers a solution to the challenge of priority inversion in real-time systems using GPGPUs, where high-priority tasks are blocked by low-priority memory transfers and kernel executions. The authors introduce a lightweight approach that enables preemptive memory copies and task executions in GPGPUs, allowing these operations to run concurrently and improving system responsiveness. Their experimental results show that the proposed system significantly reduces response times and outperforms traditional non-preemptive systems, providing a practical method for enhancing real-time performance in GPGPU-based applications.  More recently, Baek etal. comes with \cite{baek2020carss}, a portable, tagging-based cooperative scheduler and resource monitor for heterogeneous applications sharing a single hardware accelerator in a soft real-time environment. They introduce a software-based GPU activity-tagging mechanism that allows multiple client applications to run concurrently with predictable performance. Their approach, tested on both GPU and embedded platform, supports priority scheduling and does not require modifying proprietary drivers. By analyzing application-specific GPU usage patterns, they show that efficient resource sharing can improve performance without additional hardware, and their framework is extensible to other hardware accelerators. Another category is to spatially partition the accelerator during the queuing. TimeGraph \cite{kato2011timegraph} proposed a fixed-priority-based scheduling mechanism for managing GPU resources for soft real-time tasks. It aimed to control the latency and execution time of GPU-accelerated tasks by extending the Linux kernel to support deadline-aware GPU scheduling. Mangharam et al. \cite{mangharam2011anytime} explore the development of anytime algorithms for GPU architectures, specifically using the NVIDIA CUDA platform, to balance the trade-off between output quality and execution time in time-sensitive applications. They propose a PAP$^*$ algorithm that provides progressively better results with more computation time and can generate partial outputs if interrupted. The authors focus on dynamically selecting GPU resources and adjusting execution paths to optimize results within a specified deadline. A case study on a GPU-based vehicle traffic simulator, AutoMatrix, demonstrates how such algorithms can handle large-scale, real-time tasks like traffic congestion prediction.

Apart from queuing, researchers also delve into other scheduling strategy such as setting virtual deadlines. TimeWall \cite{amert2021timewall} introduces a time-partitioning framework for multicore and accelerator platforms, addressing the challenge of maintaining temporal isolation in safety-critical systems with shared accelerators. Unlike prior work, which has largely focused on uniprocessor partitioning or multiprocessor setups without accelerators, TimeWall enforces temporal isolation through a two-level scheduler that includes ``forbidden zones" to prevent accelerator access from exceeding time-slice boundaries. Previous GPU arbitration efforts, such as real-time locking protocols or driver-level modifications, have not addressed the complexities of partitioned scheduling on heterogeneous platforms. Elloit \cite{elliott2012globally} proposes a container method for accessing the GPU, targeting at predictable system performance and maximizing computing throughput. Temporal isolation is achieved in the container mechanism by allocating execution timeslot in hierarchy, making the schedulability test tight and accurate.

\subsubsection{Scheduling for Multi-Accelerator Based System}

The scheduling of multi-accelerator-based systems can be classified into two categories based on modeling and scheduling granularity. Some works focus on the task unit level, often dealing with synthetic workloads, while others operate at a near-application level, such as scheduling neural network layers.

At the task unit level, GPUSync \cite{elliott2013gpusync}, employs both fixed-priority and dynamic-priority scheduling for real-time GPU-based systems. GPUSync extends traditional real-time scheduling models to GPU-based systems, offering priority-based arbitration between tasks running on CPUs and GPUs. It uses predictable synchronization mechanisms to ensure tasks meet their timing requirements, making it one of the first works to propose fixed-priority real-time scheduling for GPU-based systems. Locking protocol \cite{elliott2013exclusion} such as k-exclusion is also well studied, which enables minimized sharing resource waiting time and maximized CPU availability.

At the near-application level, DREAM \cite{kim2024dream} introduces a scheduling framework for real-time multi-model machine learning (RTMM) workloads in accelerator-based systems. By utilizing a MapScore metric for scheduling decisions and incorporating adaptive techniques like Supernet switching and preemptive frame dropping, DREAM efficiently handles heterogeneous and unpredictable RTMM demands. The approach achieves a 32.2\% to 50\% reduction in the author-defined metrics, outperforming existing schedulers in managing complex, multi-accelerator workloads. Pegasus \cite{gupta2011pegasus}, on the other hand, coordinates scheduling for heterogeneous systems, treating accelerators as schedulable resources within a hypervisor environment. Building on prior virtualization approaches, Pegasus extends Xen's credit-based CPU scheduler to manage GPU resources using multiple policies: AccCredit (proportional GPU sharing), CoSched (simultaneous scheduling of CPU and GPU tasks), AugC (credit-boosting for CPU-scheduled VMs), and SLAF (feedback-driven adjustments for QoS). This coordination outperforms standard GPU drivers in mixed workloads, improving resource fairness and throughput. It aligns with efforts like GViM for QoS-aware GPU sharing and extends beyond traditional gang scheduling for tightly coupled CPU-GPU tasks.


\newcommand{\dlcell}[2]{\begin{tabular}[c]{@{}l@{}}{#1}\\{#2}\end{tabular}}
\newcommand{\tlcell}[3]{\begin{tabular}[c]{@{}l@{}}{#1}\\{#2}\\{#3}\end{tabular}}

{\renewcommand{\arraystretch}{2.0}
\begin{tiny}
\begin{table*}[]
\caption{Summary of soft real-time scheduling on accelerator-based heterogeneous computing.}
\label{tab:hrt}
\begin{tabular}{|l|l|l|l|l|}
\hline
\textbf{Work} & \textbf{Model} & \textbf{Objective} & \textbf{Accelerator Features}  & \textbf{Description}                                                                  \\ \hline 
PKM \cite{basaran2012supporting}  & RIM             & Timing Critical                & Temporal Division                          & Framework for preemptive GPU executions and data copies        \\ \hline
TimeGraph\cite{kato2011timegraph} & Self-suspension & Timing Critical                & Spatial Partitioning                      & Device-driver level scheduling with 2 policies   \\ \hline
GPUSync\cite{elliott2013gpusync}  & Self-suspension & Timing Critical                & Synchronizing                          & Flexible, predictable and parallel multi-accelerator system  \\ \hline
Elliott\cite{elliott2012globally} et al. & Self-suspension & Timing Critical         & Container Access                         & Analysis for global GPU accessing in multi-CPU system \\ \hline
Elliott\cite{elliott2013exclusion} et al. & Self-suspension & Timing Critical        & Locking
           & Optimal k-exclusion locking protocol for multi-GPU \\ \hline
Pegasus\cite{gupta2011pegasus}    & Not specified   & Timing Critical                & Virtualization                                    & Hypervisor level scheduling across multiple VMs    \\ \hline
PAP$^*$\cite{mangharam2011anytime}& Conditional DAG & Timing Critical                & Queuing                                           & Computing path selection to provide anytime output \\ \hline
CARSS\cite{baek2020carss}         & RIM             & Timing Critical                & Semi-temporal Division                          & Scheduling applications running in parallel on one GPU \\ \hline
TimeWall\cite{amert2021timewall}  & RIM             & Timing Critical                & Temporal Division                          & Framework for time isolation on multi-core+accelerator \\ \hline
DREAM\cite{kim2024dream}          & Not specified   & Timing Critical                & Queuing by Scoring                           & Scheduler for dynamic multi-model ML workloads   \\ \hline
sBEET\cite{wang2021balancing}     & RIM             & Power + Timing                 & \makecell[l]{Spatial Partitioning \\ Power Gating} & Framework to balance energy and timing of GPU kernels\\ \hline
sBEET-mg\cite{wang2022towards}    & RIM             & Power + Timing                 & Spatial Partitioning                      & Extension of sBEET to multiple GPUs            \\ \hline
Maity et al.\cite{maity2022future}& DAG             & Thermal + Timing               & \makecell[l]{Spatial Partitioning\\DVFS}      & Model Predictive Control based thermal-aware scheduling \\ \hline
TherMa-MiCs\cite{safari2022thermamics}& DAG         & Thermal + Timing               & \makecell[l]{Temporal Division\\DVFS}      & Thermal-aware scheduler for mixed-critical systems  \\ \hline
\end{tabular}
\end{table*}
\end{tiny}
}

\subsection{Designs from Researchers on Soft Real-Time with Other Objectives}
\label{subsec:energy_srt}

To improve both the schedulability and the energy efficiency on CPU-GPU heterogeneous computing platform, Wang \cite{wang2021balancing} presents \textit{sBEET} a real-time energy-efficient GPU scheduler that makes scheduling decisions at runtime to optimize the energy consumption while utilizing spatial multitasking to improve real-time performance. At runtime, the \textit{sBEET} makes scheduling decisions and adjusts the partitioning of computing resources, e.g., streaming multiprocessors (SMs) in NVIDIA GPUs, based on the prediction of energy consumption calculated by the power model. By choosing the partitioning of computing resources and considering computation energy and task deadlines, \textit{sBEET} reduces deadline misses and energy consumption up to 13\% and 21\% on the NVIDIA Jetson Xavier AGX. Meanwhile, a critical and universal theorem is also proposed and further proved in this work which states that the schedule of a job set $\tau$ with spatial multitasking cannot be more energy-efficient than the schedule without spatial multitasking if the jobs in $\tau$ are linear-speedup jobs. Following this work, Wang \cite{wang2022towards} further extended the scheduling strategy to \textit{sBEET-mg}, addressing the timing and energy efficiency for heterogeneous multi-GPU systems.

Thermal compliance during real-time computing and scheduling is another critical metric focused by the researchers. TherMa-MiCs \cite{safari2022thermamics}, a thermal-aware scheduling scheme designed for fault-tolerant Mixed-Criticality Systems (MCSs). These systems integrate tasks of varying criticality levels, and the study focuses on handling the thermal challenges of such platforms, especially when using fault-tolerant techniques like N-Modular Redundancy (NMR). The key challenge addressed is maintaining both temperature and timing constraints while optimizing the quality of service (QoS) for low-criticality tasks. The proposed approach aims to balance these factors while ensuring system reliability. Experimental results show that the method not only meets temperature and timing requirements but also improves QoS for low-criticality tasks by an average of 44\%.
Maity et al. \cite{maity2022future} introduce a future-aware dynamic thermal management framework for CPU-GPU embedded platforms using Model Predictive Control (MPC). The framework predicts thermal states based on upcoming tasks, allowing it to optimize task scheduling, migration, and frequency tuning to minimize peak temperatures while meeting real-time constraints. Evaluated on an Odroid-XU4, the approach leverages OpenCL for task partitioning across CPU and GPU and reduces thermal peaks compared to traditional dynamic thermal management techniques.

\section{Real-time Scheduling and Analysis for Hard Real-Time Tasks}
\label{sec:harddeadline}

\begin{figure*}
\centering
\includegraphics[width=0.99\textwidth]{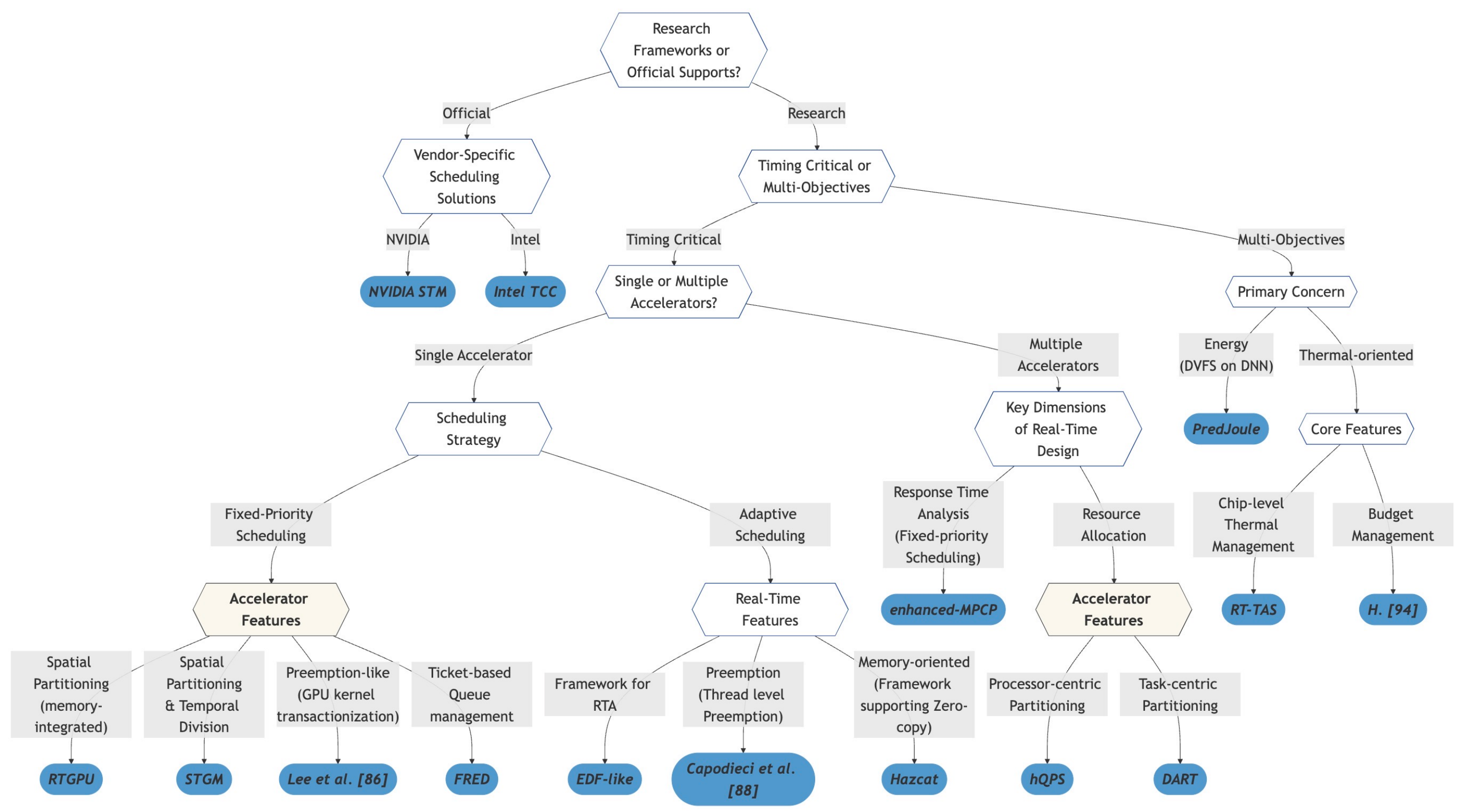}
\caption{Tree Diagram for Hard Real-Time Tasks Scheduling.}
\label{fig:hrt_classify}
\end{figure*}

This section begins by introducing solutions provided by processor vendors, followed by a primary focus on research-driven approaches to hard real-time task scheduling. Research in this area is expected to provide deadline guarantees, typically through schedulability analysis based on response time evaluation or other formal techniques. The surveyed research is mainly categorized according to its objectives, with the majority addressing timing-only requirements. A smaller portion targets multi-objective goals, such as combining timing constraints with thermal or energy considerations. Additionally, the timing-focused studies are further classified based on their target platforms, including heterogeneous architectures with either a single accelerator or multiple accelerators. Fig. ~\ref{fig:hrt_classify} presents a tree diagram that organizes the existing research on hard real-time task scheduling, while Table~\ref{tab:hrt} summarizes the key characteristics of the selected works.

\subsection{Designs from Processor Vendor}
\label{subsec:hrt_official}
Several vendors have proposed scheduling solutions tailored to their hardware platforms in the context of hard real-time systems, where tasks must meet strict deadlines without any deadline misses. This subsection highlights two representative vendor-provided scheduling approaches.

The NVIDIA System Task Manager (STM) \cite{nvidia_stm} is a scheduling framework designed to manage system-wide GPU scheduling through a computational graph-based model. STM architecture includes static and runtime components. In the offline phase, STM generates a static, time-triggered schedule based on a system-level task graph, specifying execution timing, dependencies, and resource allocation. During the runtime phase, a centralized monitor strictly enforces this schedule by dispatching tasks at predefined times. The runtime logs are leveraged for schedule optimization, allowing developers to refine the schedule offline to further improve efficiency. The static schedule ensures high predictability and minimal runtime overhead, which is essential in safety-critical systems. However, the design requires complete regeneration of the schedule whenever task characteristics are modified, limiting its adaptability in dynamic or rapidly evolving systems.

While STM focuses on static task-level scheduling to guarantee timing predictability within a tightly controlled execution environment, Intel Time Coordinated Computing (TCC) \cite{intelTCC} provides real-time assurance from a broader system-level perspective. Instead of a scheduler design, TCC defines a set of coordinated hardware and software mechanisms that collectively support time-sensitive execution across heterogeneous system components. In TCC, the timeliness is achieved by reducing latency, minimizing jitter, and enhancing determinism through features like cache reservation and time synchronization across system components.

\subsection{Designs from Researchers on Only Hard Real-Time Constraint}
\label{subsec:hrt_timingOnly}
The research on real-time scheduling with timing-critical constraints are classified into two categories based on the number of accelerators supported. For single accelerator systems, scheduling approaches are divided into fixed-priority and adaptive strategies. For multiple accelerators, the studies focus on response time analysis or resource allocation to meet real-time constraints in heterogeneous environments.

\subsubsection{Single Accelerator}
In heterogeneous systems with a single accelerator, a wide range of real-time scheduling and analysis studies adopt fixed-priority scheduling due to its analyzability and practical applicability. Within this setting, different works propose distinct methods for worst-case response time analysis or scheduling framework design, primarily differentiated by how tasks interact with the accelerator. Among the surveyed works, the employed access mechanisms can be broadly grouped into four categories: spatial partitioning, spatial partitioning with temporal division, preemption-like access, and ticket-based queue management. The following sections present representative approaches under each category.


{\renewcommand{\arraystretch}{2.0}
\begin{tiny}
\begin{table*}[]
\caption{Summary of hard real-time scheduling on accelerator-based heterogeneous computing.}
\label{tab:hrt}
\begin{tabular}{|l|l|l|l|l|}
\hline
\textbf{Work} & \textbf{Model}            & \textbf{Objective} & \textbf{Accelerator Features}  & \textbf{Description}                                    \\ \hline
SCAIR-OPA \cite{huang2015schedulability}    & Self-suspension                                & Timing Critical        & N/A                            & pure mathematical calculation on WCRT                   \\ \hline
RTGPU \cite{zou2021rtgpu}        & Self-suspension                                 & Timing Critical        & Spatial Partitioning           & Memory operation integrated in analysis                 \\ \hline
SHAPE \cite{xu2022shape}        & Self-suspension                                 & Timing Critical        & Spatial Partitioning           & Resource pool viewpoint                                 \\ \hline
STGM \cite{saha2019stgm}          & Self-suspension                                & Timing Critical        & Spatial + Temporal             & Worst-Fit Decreasing resource allocation on GPU         \\ \hline
Lee et al. \cite{lee2018gpu}      & Not specified                                  & Timing Critical        & Transactionization             & Transactionize GPU kernel into segments                 \\ \hline
EDF-like \cite{gunzel2022edf}     & Self-suspension                                 & Timing Critical        & N/A                            & Framework for EDF-like scheduling analysis              \\ \hline
FRED \cite{biondi2016framework}     & Self-suspension                                 & Timing Critical        & Partition \& global queue                            & Framework for ticket-based FPGA queue            \\ \hline

Capodieci et al. \cite{capodieci2018deadline}      & RIM                       & Timing Critical        & \makecell[l]{thread-level preemption}        & Framework for thread-level preemptive GPU        \\ \hline
Hazcat \cite{bell2023hardware}        & Task Chain                            & Timing Critical        & N/A                            & Framework supporting zero-copy                          \\ \hline
\makecell[l]{enhanced-MPCP \cite{patel2018analytical}} & Self-suspension                                & Timing Critical        & Priority queue                 & Extension of MPCP into self-suspension model            \\ \hline
hQPS   \cite{massa2021heterogeneous}         & RIM                       & Timing Critical        & Quasi-partitioning             & Partitioning processors into temporal slices \\ \hline
DART \cite{xiang2019pipelined}         & DAG                       & Timing Critical        & \makecell[l]{Semi-temporal\\partitioning}     & Decompose DNN inference into stages; pipeline           \\ \hline
PredJoule \cite{bateni2018predjoule}    & Task Chain                       & Energy + Timing    & Queue                          & DVFS; Decompose DNN inference into layers               \\ \hline
RT-TAS \cite{lee2019thermal}        & RIM                       & Thermal + Timing   & \makecell[l]{Temporal Division\\(mutex lock)} & Chip-level thermal management                           \\ \hline
Hosseinimotlagh et al. \cite{hosseinimotlagh2019thermal}          & RIM                        & Thermal + Timing   & Priority queue                 & Global thermal budget management                         \\ \hline
\end{tabular}
\end{table*}
\end{tiny}
}

RTGPU \cite{zou2021rtgpu} targets a heterogeneous platform consisting of one CPU, one memory copy engine, and a single accelerator. To mitigate the inter-task interference on the non-preemptive accelerator, RTGPU adopts spatial partitioning, assigning a fixed number of compute units (e.g., streaming multiprocessors) to each task. This isolation significantly reduces the worst-case waiting time in the response time analysis, leading to tighter schedulability bounds. The proposed method explicitly models memory transfer as a separate stage and incorporates it into the overall response time computation. The spatial partitioning can be achieved using mechanisms like Multi-Process Service (MPS) \cite{nvidia_mps} or Multi-Instance GPU (MIG) \cite{nvidia2023mig} feature. SHAPE \cite{xu2022shape} also applies spatial partitioning to eliminate task interference on the accelerator, while extending the analysis framework to a platform with multiple CPUs and one shared accelerator. Although SHAPE does not introduce new access strategy for the accelerator, its primary contribution lies in a resource pool-based analysis method, which calculates schedulability by comparing the available computing resources with the upper bounds of task-induced resource demand, with respect to time. This abstraction enables unified reasoning across processors while preserving execution predictability on the accelerator.

STGM \cite{saha2019stgm} introduces a GPU management framework that combines spatial partitioning with temporal coordination to improve schedulability under fixed-priority scheduling. It partitions GPU streaming multiprocessors (SMs) among tasks using a Worst-Fit Decreasing heuristic guided by response time analysis. When SMs are shared, FIFO-based kernel execution and priority boosting are applied to reduce interference and improve temporal isolation. Unlike hardware-based isolation mechanisms such as MIG, STGM does not enforce strict SM exclusivity. Instead, it relies on software-level control to guide task execution to designated SMs. As a result, when resource constraints or heuristic allocation lead to SM sharing, the temporal management acts as a compensatory mechanism, serializing access through FIFO-based ordering to maintain predictability.

Unlike CPUs, accelerators like GPUs generally lack native support for preemption operations, though recent research has explored methods to enable preemption-like behavior in GPUs. Lee et al. \cite{lee2018gpu} proposes a preemption-like scheduling method, where transactional kernel execution is introduced for accelerator-side execution. Here, transactionization refers to dividing GPU kernel into smaller and independent transactions. Although this does not provide real preemption, it allows high-priority tasks to access accelerator resources more quickly by minimizing the blocking time. This mechanism is further supported by a snapshot mechanism, which rolls back and restores the context when real preemption is required. 

Compared to GPUs, FPGAs offer reconfigurable hardware logic, allowing customization for specific applications but introducing challenges such as higher reconfiguration overhead and complex resource management. To address the resource contention issues caused by dynamic partial reconfiguration (DPR), FRED \cite{biondi2016framework} introduces a ticket-based task queue management mechanism. By assigning timestamps (tickets) to task requests and sorting them, the framework constructs partition-specific queues for slot scheduling within FPGA partitions and a global queue for managing access to the FPGA Reconfiguration Interface, FRI. The authors further introduce hardware-software co-design, including user-level API support, FPGA partitioning and slot management, and direct memory access to improve reconfiguration predictability, and achieve satisfactory speedup.


In contrast to fixed-priority scheduling, some works explore adaptive scheduling strategies that dynamically determine task execution priorities or resource access based on runtime conditions or task characteristics. These approaches differ significantly in their objectives and mechanisms, reflecting the diversity of adaptive scheduling in heterogeneous systems. EDF-like \cite{gunzel2022edf} targets a general suspension-aware schedulability analysis framework, and proposes a response time analysis method for a class of job-level fixed-priority scheduling strategies, where job-level refers to making scheduling decisions based on each individual release of a task. The proposed analysis supports both constrained and arbitrary-deadline task sets, and accommodates a wide range of suspension models, including segmented, dynamic, and hybrid behaviors. While the framework is analytically general, it is fundamentally built upon a uniprocessor self-suspension model. Although the self-suspension abstraction is also widely used to model accelerator segments in heterogeneous systems, some of the analytical techniques in EDF-like may not directly carry over to heterogeneous environments.

Capodieci et al. \cite{capodieci2018deadline} proposed a deadline-based scheduling framework integrating EDF and Constant Bandwidth Server (CBS) strategies to provide GPU preemption support. By leveraging NVIDIA Pascal GPU's thread-level preemption, the scheduler can dynamically interrupt low-priority tasks and have high-priority tasks to timely access GPU resources with minimal delay. CBS provides temporal isolation by enforcing per-task execution budgets, enabling safe sharing of GPU time among tasks with varying urgency levels. Hazcat \cite{bell2023hardware}, on the other hand, shifts focus from scheduling policies to runtime system optimization. It introduces a zero-copy memory management mechanism to reduce memory transfer latency and unpredictability, thereby enhancing the performance and determinism of real-time systems. Zero-copy \cite{che2011dymaxion} refers to a data transfer technique that avoids redundant copying of data between different memory locations, such as between host memory and device memory. Instead, it enables direct access to data in its original memory location, minimizing overhead and improving efficiency. 

\subsubsection{Multiple Accelerators}
In real-time systems equipped with multiple accelerators, existing works generally fall into two broad categories. The first continues the line of response time analysis, extending the analytical techniques developed for single-accelerator settings to multi-accelerator environments. These studies typically focus on schedulability guarantees under fixed task-to-accelerator mappings or statically partitioned workloads. The second category addresses resource allocation and mapping problems, which arise due to the increased flexibility and complexity of multi-accelerator platforms. In this context, the research focus shifts from purely satisfying hard real-time constraints to optimizing system-level objectives such as throughput and latency, while still ensuring deadline compliance.

For response time analysis in multi-accelerator environments, existing approaches largely extend techniques developed for single accelerator systems. Most analyses adopt a CPU-centered perspective, as CPU supports preemption and thus reduces difficulty and pessimism in the analysis. Enhanced-MPCP \cite{patel2018analytical} exemplifies this direction by introducing improved blocking time analysis for self-suspending tasks under the Multiprocessor Priority Ceiling Protocol (MPCP).

In contrast, another type of work addresses the task-to-accelerator allocation problem under timing constraints. These approaches aim to balance system throughput and latency while preserving schedulability. hQPS \cite{massa2021heterogeneous} proposes a quasi-partitioned scheduling method, which involves a spatial partitioning-like strategy, dividing accelerator executions into small pieces to reduce blocking and cost of task migration. These migrations serve as a form of temporal load balancing, redistributing tasks from overloaded accelerators to lightly loaded ones. hQPS employs Mixed-Integer Linear Programming (MILP) to optimize the task-to-processor assignments and utilizes a quasi-partitioned scheduler during runtime to further balance task loads across accelerators. DART \cite{xiang2019pipelined}, similarly, introduced a semi-temporal division strategy into the scheduling of multi-accelerator cases. Specifically, it leverages a pipeline structure, where DNN inference tasks are decomposed into independent computation slices, referred to as stages, and executed in a pipeline fashion. While the pipeline itself is a temporal scheduling approach, DART integrates a task-to-accelerator assignment process, making it a task-centric division strategy. Both hQPS and DART explore different strategies for mapping tasks to accelerators, with hQPS emphasizing spatial fragmentation and load balancing, while DART adopting a structured, stage-wise pipeline mapping.

\subsection{Designs from Researchers on Hard Real-Time with Other Objectives}
Real-time scheduling with multi-objectives extends traditional hard real-time system design by incorporating additional optimization goals—most notably, energy efficiency and thermal safety—while still preserving strict timing guarantees. In contrast to the soft real-time context discussed in Section~\ref{subsec:energy_srt}, these works maintain a strong emphasis on hard real-time properties, such as worst-case response time analysis and deadline satisfaction. The following studies represent different approaches to integrating energy- or thermal-aware strategies into real-time scheduling frameworks.

PredJoule \cite{bateni2018predjoule} addresses the energy optimization problem for real-time DNN inference by introducing a layer-aware Dynamic Voltage and Frequency Scaling (DVFS) strategy. Specifically, it dynamically adjusts the voltage and frequency for each layer based on the its computational and energy consumption characteristics, guided by a feedback-based progress tracker and a learning-based controller. Although PredJoule runs on a heterogeneous platform, it does not explicitly present scheduling decisions for such CPU-GPU environments, such as mapping zhDNN stages to different processing units. This reflects a trend in multi-objective real-time research, where energy control is often decoupled from scheduling decisions. Nevertheless, PredJoule’s integration of runtime adaptation and uncertainty-aware control offers valuable insight into energy-efficient execution under real-time constraints.

In terms of thermal safety, RT-TAS \cite{lee2019thermal} addresses the challenge of managing thermal hotspots in real-time systems by employing a thermal-balanced task allocation approach. It proposes a Worst-Fit Decreasing (WFD) algorithm that dynamically allocates tasks based on their power consumption and thermal coupling characteristics, assigning high-heat tasks to cooler regions of the chip to mitigate temperature fluctuations. This strategy is particularly focused on maintaining chip-level thermal safety, ensuring that task execution does not exceed thermal limits while still meeting real-time deadlines. In contrast to PredJoule, RT-TAS effectively integrates thermal management with scheduling decision, providing a robust solution for ensuring both thermal safety and timing predictability in complex, multi-task environments. Similarly, Hosseinimotlagh et al. \cite{hosseinimotlagh2019thermal} introduces a thermal-aware server-level framework for global task management, which combines server budget management with task scheduling. The scheduling and resource management strategy in this paper is similar to mixed-criticality approaches, where higher-priority tasks are allocated more resources, ensuring both timing guarantees and thermal safety.

In summary, research on hard real-time task scheduling with a \textit{timing-only} objective is predominantly focused on single-accelerator environments. These works tend to leverage the Multi-Segment Self-Suspension model for response time analysis to determine whether tasks may experience deadline misses. In contrast, in multi-accelerator environments, the emphasis shifts more towards task-to-accelerator allocation strategies, leading to a relative de-emphasis on response time analysis. For the \textit{multi-objective} category, studies on hard real-time tasks are fewer compared to those on soft real-time tasks, as hard task constraints are often relaxed when addressing additional objectives, such as energy efficiency or thermal safety.

\section{Application/Scenario Driven Real-Time Scheduling}
\label{sec:application}

\begin{figure*}
\centering
\includegraphics[width=0.99\textwidth]{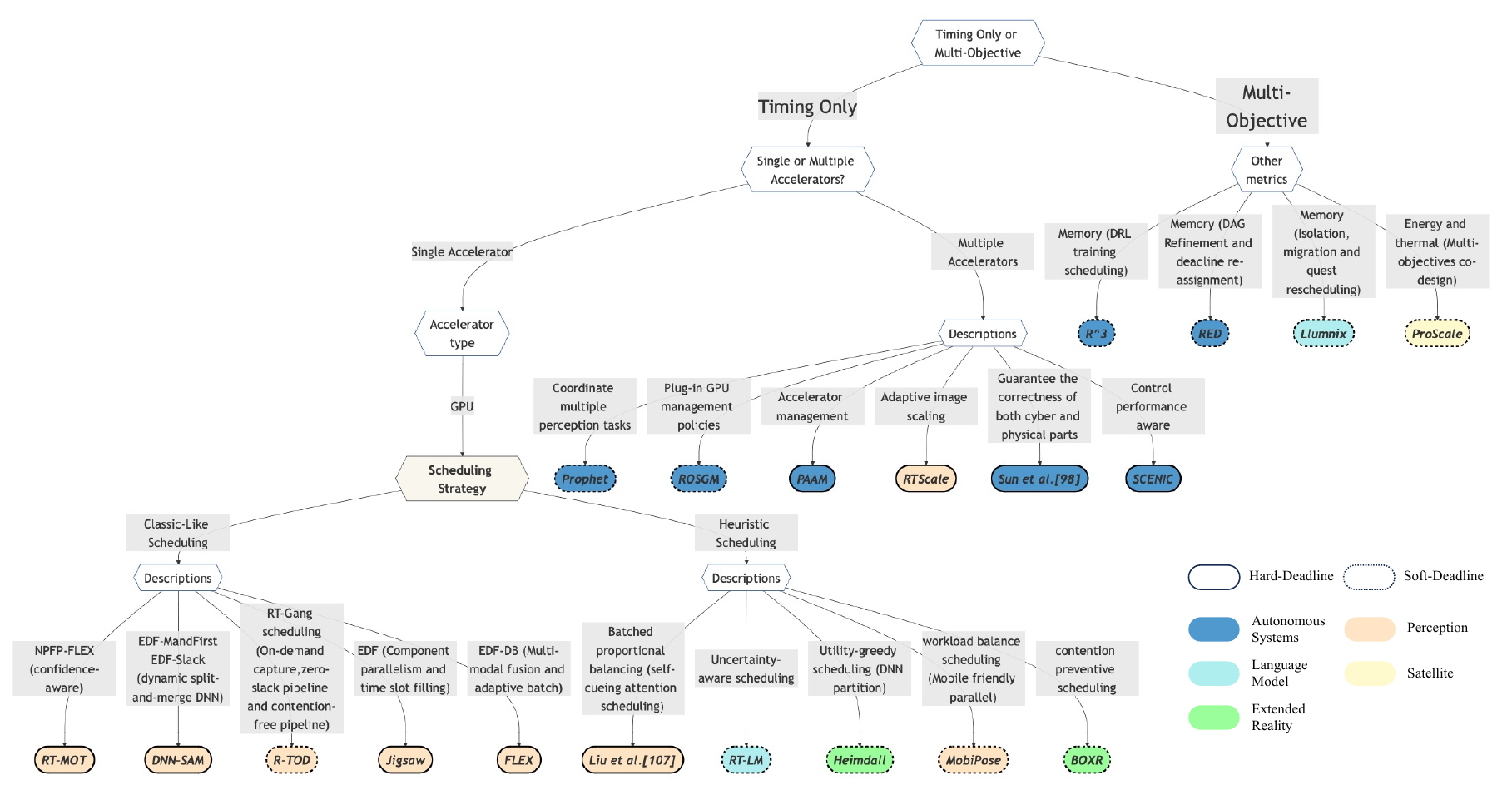}
\caption{Tree Diagram for Applications.}
\label{fig:app_classify}
\end{figure*}

In this section, we explore applications that require real-time computing on heterogeneous platforms, as well as application-driven scheduling approaches designed on accelerator-based heterogeneous architectures. While some applications demand strict guarantees for meeting execution deadlines, others prioritize optimizing execution speed to enhance overall performance. Real-time applications are typically developed within comprehensive frameworks that address not only timing constraints but also additional objectives such as energy efficiency and thermal management. These considerations are crucial to ensuring the reliability and sustainability of real-time systems operating in diverse and often resource-constrained environments.

\begin{tiny}
\begin{table*}[]
\caption{Summary of Applications.}
\begin{tabular}{|l|l|l|l|l|l|l|}
\hline
\textbf{\begin{tabular}[c]{@{}l@{}}Application \\ Field\end{tabular}} & \textbf{Work} & \textbf{Task Model} & \textbf{\textbf{\begin{tabular}[c]{@{}l@{}}Scheduling \\ Type  \end{tabular}}}& \textbf{\textbf{\begin{tabular}[c]{@{}l@{}}Scheduling \\ Objective  \end{tabular}}} & \textbf{\begin{tabular}[c]{@{}l@{}}Accelerator \\ Type (Features) \end{tabular}} & \textbf{\begin{tabular}[c]{@{}l@{}}Descriptions\end{tabular}}\\ \hline
\multirow{16}{*}{\textbf{\begin{tabular}[c]{@{}l@{}}Autonomous \\ Systems\end{tabular}}}  
& $R^3$~\cite{li2023mathrm}  & 
RIM
& Soft Real-Time & \dlcell{Multi-objectives}{(Memory-Driven)}  & Single GPU & \dlcell{DRL training}{scheduling}\\ \cline{2-7} 
& Prophet~\cite{liu2022prophet}  & 
RIM
& Soft Real-Time & \tlcell{Timing only}{(Classical-Like:}{Linux patch)}  & Multiple GPUs & \dlcell{Coordinate multiple}{perception tasks}\\ \cline{2-7} 
& Sun et al.~\cite{10155700}  & 
DAG
& Hard Real-Time & \tlcell{Timing only}{(Classical-Like:}{EDF)}  & Multiple GPUs & \dlcell{Guarantee the correctness of }{both cyber and physical parts}\\ \cline{2-7} 
& SCENIC~\cite{chen2024scenic}  & 
MSSS
& Hard Real-Time & \tlcell{Timing only}{(Heuristic:}{control performance)}  & Multiple GPUs & \dlcell{Control performance}{aware}\\ \cline{2-7} 
& Red~\cite{li2023red}  & DAG & Soft Real-Time & \dlcell{Multi-objectives}{(Memory-Driven)}  & Single GPU & \dlcell{DAG Refinement and}{deadline re-assignment}\\ \cline{2-7} 
& ROSGM~\cite{10155690}  & MSSS & Soft Real-Time & \tlcell{Timing only}{(Classic-Like:}{FIFO)} & Multiple GPUs & \dlcell{Plug-in GPU}{management policies} \\ \cline{2-7} 
& PAAM~\cite{10568046} & Task chain & Hard Real-Time & \tlcell{Timing only}{(Heuristic:}{criticality-as-priority)}  & \dlcell{Multiple GPUs}{and TPUs} & \dlcell{Accelerator}{management}\\ \cline{2-7} 
\hline
\multirow{22}{*}{\textbf{Perception}} 
& RTScale~\cite{heo_et_al:LIPIcs.ECRTS.2022.2}  & 
RIM
& Hard Real-Time & \tlcell{Timing only}{(Heuristic:}{sensitivity prediction)} & Multiple GPUs & Adaptive image scaling\\ \cline{2-7} 
& RT-MOT~\cite{9984748} & RIM & Hard Real-Time & \tlcell{Timing only}{(Classic-Like:}{$NPFP^{flex}$)} & Single GPU & \dlcell{confidence-aware}{real-time scheduling}\\ \cline{2-7} 
& DNN-SAM~\cite{9804671} & RIM & Hard Real-Time & \tlcell{Timing only}{(Classic-Like:}{EDF-MandFirst,EDF-Slack)} & Single GPU & \dlcell{dynamic split-and-merge DNN}{execution and scheduling}\\ \cline{2-7} 
& R-TOD~\cite{9355528} & RIM & Soft Real-Time & \tlcell{Timing only}{(Classic-Like:}{RT-Gang scheduling)} & Single GPU & \tlcell{On-demand capture,}{zero-slack pipeline and}{contention-free pipeline}\\ \cline{2-7} 
& Liu et al.~\cite{liu2022self} & RIM & Hard Real-Time & \tlcell{Timing only}{(Heuristic: Batched}{proportional balancing)} & Single GPU & \dlcell{Self-cueing attention}{scheduling}\\ \cline{2-7} 
& MobiPose~\cite{zhang2020mobipose} & 
RIM
& Soft Real-Time & \tlcell{Timing only}{(Heuristic:}{workload balance)} & Single GPU & Mobile friendly parallel\\ \cline{2-7} 
& Jigsaw~\cite{sun2024jigsaw}  & 
RIM
& Hard Real-Time
& \tlcell{Timing only}{(Classic-Like:}{EDF)} & \dlcell{Single GPU}{(Preemptive)} & \dlcell{Component parallelism}{and time slot filling}\\ \cline{2-7} 
& Flex~\cite{xu2024flex}  & DAG & Hard Real-Time & \tlcell{Timing only}{(Classic-Like:}{EDF-DB)} & Single GPU & \dlcell{Multi-modal fusion}{and adaptive batch}\\ \cline{2-7} 
\hline
\multirow{3}{*}{\textbf{\begin{tabular}[c]{@{}l@{}}Language \\ Model\end{tabular}}}  
& RT-LM~\cite{10405961}  & RIM & Soft Real-Time & \tlcell{Timing only}{(Heuristic:}{uncertainty-based)}   & Single GPU & \dlcell{Uncertainty-aware}{scheduling}\\ \cline{2-7} 
& Llumnix~\cite{298685}  & RIM & Soft Real-Time & \dlcell{Multi-objectives}{(Memory)} & Multiple GPUs & \dlcell{Isolation, migration and}{quest rescheduling}\\ \cline{2-7} 
\hline
\multirow{1}{*}{\textbf{Satellite}} 
& ProScale~\cite{li2024satellite}  & 
RIM
& Soft Real-Time
& \dlcell{Multi-objectives}{(Energy and thermal)} & 
Single NPU
& \dlcell{Multi-objectives}{co-design} \\ \cline{2-7} 
\hline
\multirow{4}{*}{\textbf{\dlcell{Extended}{Reality}}} 
& BOXR~\cite{Zhang2024boxr}  & 
MSSS
& Soft Real-Time 
& \tlcell{Timing only}{(Heuristic:}{contention-preventive)} & single GPU &   \tlcell{Motion-driven visual inertial}{Odometer and scene-dependent}{foveated rendering} \\ \cline{2-7} 
& Heimdall~\cite{yi2020heimdall}  & 
RIM
& Soft Real-Time & \tlcell{Timing only}{(Heuristic:}{utility-greedy)} & Single GPU & DNN partition\\ \cline{2-7} 
\hline
\end{tabular}
\end{table*}
\end{tiny}

\subsection{Autonomous Systems}
Recently, the research of autonomous systems in the field of real-time has been thoroughly studied in various aspects. An autonomous system refers to a self-governing entity capable of making decisions and performing tasks independently without human intervention. It leverages advanced technologies, such as artificial intelligence, machine learning, and sensor integration, to perceive its environment, analyze data, and execute actions. Autonomous systems are widely applied in fields such as autonomous vehicles, robotics, aerospace, and industrial automation, where real-time decision-making and adaptability are critical for performance and safety. 

Autonomous systems typically execute multiple tasks concurrently, often with intricate dependencies or competition for limited resources. These interactions significantly complicate the scheduling design, requiring sophisticated strategies to ensure efficient and reliable operation. In the context of autonomous vehicles, Liu et al.~\cite{liu2022prophet} conducted a comprehensive empirical study and identified two key insights that address these challenges. Based on their findings, they proposed Prophet, a framework specifically designed to tackle the issue of deep neural network (DNN) inference time variations. Prophet adopts a two-step approach: first, it predicts the time variations of individual DNN models to improve timing accuracy; second, it coordinates the inference of multiple DNN models to minimize fusion time variations, thereby enhancing overall system performance. This dual-layered approach effectively mitigates unpredictable delays, ensuring more robust and responsive behavior in autonomous systems.

A fundamental challenge in autonomous driving (AD) systems lies in the oversimplification of task dependencies, where inherent data dependencies are often sacrificed and not explicitly enforced as precedence constraints. As a result, complex data dependencies can emerge between tasks with varying activation rates, making it extremely difficult to analyze the real-time behavior of these systems. Sun et al.~\cite{10155700} introduce a novel timing analysis framework that ensures the correctness of both the cyber and physical components of the AD system. Within this framework, the "design-analysis-redesign" process is automated, allowing iterative refinement of the system. More importantly, failures identified in the current design process are leveraged to guide the redesign process, facilitating the development of a more efficient and systematic design methodology for AD systems.

SCENIC~\cite{chen2024scenic} introduces a novel end-to-end co-design framework that integrates capability analysis and scheduling to efficiently design and execute intelligent control tasks. The approach begins by defining a control capability function, which establishes a relationship between control performance, controller complexity, computational requirements, and the physical properties of the system. Building on this foundation, SCENIC optimizes algorithmic capability and resource allocation across both offline design and real-time execution stages. Finally, a case study on drone control using Microsoft AirSim demonstrates the superiority of SCENIC over state-of-the-art design methods, achieving improved performance and efficiency.
Red~\cite{li2023red} is a comprehensive framework designed for multi-task deep neural network (DNN) inference on resource-constrained robotic systems, enabling adaptive navigation of Robotic Environmental Dynamics under real-time constraints. At its core, RED features a deadline-driven scheduler with an intermediate deadline assignment policy, capable of managing dynamic workloads and asynchronous inference in unpredictable environments. Additionally, RED effectively supports the deployment of MIMONet (multi-input multi-output neural networks), overcoming memory limitations and leveraging the unique weight-sharing architecture of MIMONet. Through an innovative workload refinement and reconstruction process, RED ensures seamless compatibility with MIMONet while optimizing overall efficiency.

The Robot Operating System (ROS) is a versatile framework for building robot applications, offering tools, libraries, and conventions for tasks like perception, control, and communication. Despite its modularity and scalability, ROS was not initially designed for real-time performance. Real-time scheduling is crucial for ensuring deterministic task execution, particularly in latency-sensitive and precision-critical applications like autonomous driving or robotic surgery. Extensions such as ROS 2 enhance ROS with real-time capabilities, enabling efficient resource allocation and task prioritization to ensure responsiveness and system reliability.
ROSGM~\cite{10155690} introduces a plug-in mechanism that allows seamless integration of custom GPU management policies and supports dynamic switching between policies at runtime. Additionally, ROSGM enables dynamic loading and unloading of ROS 2 tasks within the same running application, providing efficient task management. By employing asynchronous GPU request submission and optimized GPU management strategies, ROSGM significantly improves the performance of ROS 2 applications. Furthermore, the flexibility of its plug-in policies ensures compatibility with the varying demands of different applications, as well as the ability to adapt to changes within a single application across various scenarios.
PAAM~\cite{10568046} proposed by Daniel et al. introduces an accelerator management server that operates as a standalone executor within the ROS 2 application layer, offering accelerator access as a service to clients. It focuses on the challenge of priority inversion and unbounded blocking, poor accelerator resource utilization and disparity in chain and executor priorities in ROS 2 ecosystem. By mitigating these issues, PAAM enhances the efficiency and predictability of accelerator usage in ROS 2-based systems, enabling more robust performance in time-sensitive robotic applications.

Reinforcement learning (RL) is a core research methodology in autonomous systems, enabling agents to learn optimal decision-making policies through interactions with their environment. In real-time systems, RL plays a crucial role by allowing autonomous agents to adapt dynamically to changing conditions and uncertainties while meeting stringent time constraints. 
As a state-of-the-art solution, $R^3$~\cite{li2023mathrm} is meticulously designed to guarantee timing predictability during the execution of deep reinforcement learning (DRL) training workloads on GPU-enabled autonomous embedded systems. By harnessing a thorough understanding of DRL workload characteristics and integrating real-time system feedback, $R^3$ achieves a seamless balance between timing precision and algorithmic performance. Moreover, it excels in meeting stringent memory constraints, ensuring efficient resource utilization without compromising performance. This holistic approach positions $R^3$ as a robust framework for addressing the unique challenges of DRL in real-time, resource-limited environments.

\subsection{Perception}
In recent years, perception technologies have advanced significantly, driven by breakthroughs in deep learning, sensor fusion, and edge computing. Modern systems now process high-dimensional data from cameras, LiDAR, and radar for precise object detection, tracking, and scene understanding in applications such as autonomous vehicles, robotics, and augmented reality. However, meeting real-time constraints remains a challenge, as processing delays can jeopardize decision-making and safety. Real-time scheduling is essential to allocate computational resources efficiently, prioritize critical tasks, and ensure low-latency execution, enabling reliable performance in dynamic, time-sensitive environments. 

RTScale~\cite{heo_et_al:LIPIcs.ECRTS.2022.2} is a novel framework designed to achieve real-time object detection by adaptively scaling images while minimizing accuracy degradation. The key insight driving RTScale is the observation that different images exhibit varying levels of sensitivity to scaling, which directly affects object detection accuracy. Leveraging this observation, RTScale dynamically determines the optimal scaling factor for images from multiple input streams, balancing both their scale sensitivity and the real-time performance constraints. Furthermore, RTScale enhances existing object detection models by incorporating a lightweight sensitivity inference module, consisting of a few additional layers, which efficiently predicts the sensitivity of each image to scaling. This approach ensures compatibility with existing detectors while significantly improving their adaptability and real-time performance.

Real-time multi-object tracking (MOT) is crucial for applications like autonomous driving, where timely execution and accuracy are essential. Traditional MOT systems, focused on maximizing tracking accuracy and FPS, struggle to meet the strict timing requirements of resource-constrained platforms. Existing methods overlook the complexities of multi-camera systems. RT-MOT~\cite{9984748} addresses these challenges by introducing a confidence-aware real-time scheduling framework for MOT tasks. Unlike prior approaches, it dynamically balances the trade-off between execution time and tracking accuracy by leveraging a redefined notion of object confidence, which predicts tracking accuracy variations under different workload configurations. Through a novel non-preemptive fixed-priority scheduling algorithm (NPFPflex), RT-MOT guarantees timing constraints offline while optimizing tracking accuracy at runtime. 

DNN-SAM~\cite{9804671}, a dynamic split-and-merge execution and scheduling framework for deep neural networks (DNNs) which transparently decomposes an original DNN into two sub-tasks and uses a lightweight real-time scheduler to prioritize mandatory sub-tasks over optional ones, dynamically adjusting the scale of optional sub-tasks. It is specifically designed to meet the unique requirements of real-time DNN-based object detection in autonomous vehicles, providing varying detection quality for image regions with different criticality levels while ensuring all timing constraints are met.

R-TOD~\cite{9355528} is inspired by the observation that many state-of-the-art real-time object detectors exhibit unexpectedly large end-to-end time lags, despite achieving high frame rates. To address this issue, R-TOD provides a comprehensive understanding of the end-to-end delay in object detection systems, with a specific focus on Darknet YOLO. Building on this analysis, R-TOD is composed of three optimization techniques: (i) on-demand capture to minimize unnecessary processing delays, (ii) a zero-slack pipeline to streamline operations for maximum efficiency, and (iii) a contention-free pipeline to eliminate resource conflicts and improve system performance.

Liu et al.~\cite{liu2022self} propose a self-cueing attention scheduling framework designed to optimize the efficiency of visual machine perception on resource-constrained embedded platforms, aiming to minimize location error while maintaining recall. The framework incorporates a scheduling algorithm with a theoretically proven approximation ratio for reducing maximum location uncertainty, which is implemented on an NVIDIA Jetson Xavier board. This work advances the field of attention scheduling by enabling AI-based perception pipelines to selectively process data at the subframe level, aligning with tracking and safety requirements, all without relying on external cues.

MobiPose~\cite{zhang2020mobipose} is a real-time multi-person pose estimation (PE) system optimized for mobile devices, capable of estimating human poses from live video captured by mobile cameras. To address the high computational demands of multi-person PE, MobiPose employs a motion-vector-based method to efficiently track human proposals across frames, significantly reducing redundant computations. It also features a lightweight, mobile-friendly pose estimation model that balances low latency with sufficient accuracy and utilizes an efficient parallel processing engine to maximize resource utilization. A prototype of MobiPose has been successfully implemented on multiple commodity Android devices, demonstrating its capability for real-time applications.


Bird's Eye View (BEV) is a multi-modal multi-view perception technique that projects sensor data, such as camera or LiDAR inputs, into a top-down 2D representation of the environment. BEV is widely used in autonomous driving for tasks such as object detection, lane tracking, and motion planning, as it provides a comprehensive spatial understanding of the surroundings. Real-time scheduling is essential in BEV systems to process large volumes of sensor data efficiently, ensuring timely updates of the BEV map. This is critical for maintaining low-latency decision-making and enabling smooth, responsive operations in dynamic and safety-critical scenarios.
Sun et al.~\cite{sun2024jigsaw} propose Jigsaw, a task management framework designed to deploy BEV-centric perception workloads on dual-SoC platforms while meeting real-time requirements.  Jigsaw introduces two key mechanisms: first, it leverages component parallelism to minimize BEV model latency by optimizing task execution across the platform. Second, it employs a timeslot-filling scheduling strategy for Perspective-View (PV) models, ensuring predictable latency and maintaining timing guarantees. This framework effectively balances performance and predictability, making it well-suited for real-time perception tasks such as BEV.
Xu et al.~\cite{xu2024flex} introduce FLEX, a scheduling framework designed for multi-modal, multi-view perception systems operating on resource-constrained embedded platforms with onboard GPUs. FLEX integrates an elastic multi-modal fusion strategy and an adaptive batch scheduling algorithm within a context-aware scheduling principle. This approach intelligently allocates limited computing resources to critical spatial views with higher object densities, ensuring efficient resource utilization and improved perception performance in dynamic environments
~\cite{9586231}.

\subsection{Language Model}
Language models are AI systems that process and generate human language by analyzing patterns in large-scale textual data. They underpin applications like chatbots, virtual assistants, machine translation, text summarization, and sentiment analysis. Advanced models such as GPT and BERT drive innovation across industries, including customer service, content creation, and healthcare. Real-time scheduling is vital for deploying language models in latency-sensitive applications like conversational AI and real-time translation. It ensures low-latency processing and efficient resource allocation, enabling timely responses and optimal performance, especially on resource-constrained devices like mobile or edge platforms.
Li et al. propose RT-LM~\cite{10405961}, an uncertainty-aware resource management ecosystem for real-time on-device language models (LMs). The framework quantitatively reveals how input uncertainties, well-established in the NLP community, negatively impact latency by significantly increasing the output length (i.e., the number of generated tokens). Building on this insight, it introduces a lightweight runtime method to predict output length by correlating it with a comprehensive set of input uncertainties. Furthermore, RT-LM integrates this uncertainty quantification into a system-level scheduler to optimize performance through uncertainty-aware prioritization, dynamic consolidation, and strategic CPU core utilization.
Llumnix~\cite{298685} envisions serving Large Language Models (LLMs) akin to Unix systems. This concept stems from the observation that LLMs and modern operating systems share core characteristics, such as universality, multi-tenancy, and dynamism, which lead to similar requirements and challenges. It advances this vision by applying established OS principles to LLM serving. Key contributions include defining classic abstractions like isolation and priorities within the context of LLMs, implementing "context switching" via inference request migration, and enabling dynamic request rescheduling leveraging this migration. Together, these innovations allow Llumnix to achieve lower latency, improved cost efficiency, and support for differentiated Service Level Objectives (SLOs), paving the way for a new paradigm in LLM serving.

\subsection{Satellite}
Satellite positioning, such as GPS, determines device locations using signals from a satellite network and is widely applied in navigation, satellite-aided driving, and geospatial mapping. In satellite-aided driving, it supports accurate localization, route planning, and real-time traffic monitoring. Beyond transportation, satellite data is essential for agriculture, disaster management, and environmental monitoring.
Real-time scheduling is vital in these systems to process satellite signals, fuse sensor data (e.g., IMUs or cameras), and adapt to dynamic environments under strict timing constraints. It ensures low-latency processing, precise data synchronization, and reliable performance, critical for the safety and efficiency of satellite-based applications.
ProScale~\cite{li2024satellite}, a lightweight and application-aware power management and thermal control system, sheds light on the critical impact of energy and thermal characteristics on computing performance in SmallSats, which arise from the alternating power supply between solar panels and batteries, as well as the limited onboard heat dissipation capabilities. ProScale is designed to optimize computing efficiency while ensuring strict adherence to battery discharge constraints and thermal limits, delivering significant improvements without compromising system reliability or safety.

\subsection{Extended Reality}
EXtended Reality (XR) is an umbrella term that encompasses Virtual Reality (VR), Augmented Reality (AR), and Mixed Reality (MR), representing immersive technologies that blend virtual and physical environments. XR leverages hardware such as head-mounted displays (HMDs), cameras, and sensors, combined with rendering engines and computational algorithms, to create interactive and engaging user experiences.
Real-time scheduling is essential for XR systems to deliver smooth and responsive user experiences. XR applications involve complex tasks such as rendering, sensor processing, and motion tracking, all requiring strict timing. Efficient scheduling reduces delays, avoids resource contention, and ensures synchronization between virtual and physical elements. This guarantees low-latency interactions, high-quality visuals, and precise motion tracking, critical for immersion and usability in XR.
BOXR~\cite{Zhang2024boxr}, a framework for optimizing body and head motion delays in eXtended Reality (XR) systems, addresses the challenges of co-optimizing motion latencies. It introduces C2D (Computation-to-Display delay) based on body motion delays and uses a contention-preventive scheduling policy to prevent conflicts between rendering and reprojection tasks. An on-demand IMU interface (IMUi) minimizes wasted computations during IMU processing, achieving low M2D (Motion-to-Display) and C2D latencies by efficiently managing task sequences.
BOXR also features a motion-driven visual-inertial odometer (VIO) that dynamically adapts feature extraction to motion dynamics, using an error-bounding method to correct positional inaccuracies and stay within time budgets. Additionally, Scene-Dependent Foveated Rendering adjusts the foveation area based on scene complexity, balancing high frame quality and rendering times, while optimizing system performance in XR environments.
Heimdall~\cite{yi2020heimdall} is a mobile GPU coordination platform specifically designed for emerging augmented reality (AR) applications. To effectively manage the simultaneous execution of multi-DNN and rendering tasks, Heimdall introduces a Preemption-Enabling DNN Analyzer, which partitions deep neural networks (DNNs) into smaller execution units. This enables fine-grained GPU time-sharing while maintaining minimal latency overhead for DNN inference, ensuring smooth performance. Additionally, Heimdall features a Pseudo-Preemptive GPU Coordinator, which dynamically prioritizes and schedules multi-DNN and rendering tasks across both GPU and CPU resources. This flexible coordination ensures that the platform meets the stringent performance and responsiveness requirements of AR applications, delivering a seamless user experience even under complex computational loads.

\section{Discussion and Closing Remarks}
\label{sec:remark}
Finally, we conclude this survey by presenting the challenge and open questions for real-time scheduling on accelerator-based heterogeneous architectures in this section.

\subsection{Types and Numbers of Processor Cores} 
Many real-time scheduling approaches are designed for heterogeneous architectures composed of two types of computing units—typically combinations such as CPU-GPU or CPU-FPGA interconnected via a data bus. For instance, Wang et al. \cite{wang2022towards} focused on heterogeneous systems featuring multiple GPU types. However, relatively few studies have explored architectures that integrate a broader range of processor cores—such as systems that simultaneously incorporate CPUs, GPUs, FPGAs, and other specialized accelerators. This gap is partly due to the fact that commercially available platforms from major technology companies like NVIDIA, AMD, and Xilinx generally include only two types of processors: CPUs and either GPUs or FPGAs.

In contrast, some advanced industrial-grade heterogeneous platforms support a wider variety of processing elements. For example, NVIDIA PX2 and Pegasus \cite{maity2021chauffeur} integrate up to four types of processors; EyeQ \cite{Mobileye} includes five; and Jacinto (6th or 7th Gen) \cite{venkatasubramanian20202} features more than six. Despite the increasing complexity and capability of these platforms, they are primarily developed for industrial applications, and there remains a scarcity of published research offering real-time scheduling solutions specifically tailored to them.

As demonstrated in prior work\cite{zou2021rtgpu}, integrating the architectures with more types of processors, increasing the diversity of processor types, significantly amplifies pessimism in response time analysis and schedulability testing. Therefore, developing scheduling algorithms and response time analysis techniques or extending existing ones to support architectures with multiple types of processors is essential.

\subsection{Memory Bus and Data Copy Overheads}
Unlike conventional homogeneous architectures, where the response time of a task is primarily determined by its execution time on CPU cores, accelerator-based heterogeneous architectures introduce a significant overhead due to data movement across processor memories via the memory bus \cite{mendoza2020imgprocess, xu2019resnetinf, singh2020rnnt, do2022cudasample}. With the adoption of zero-copy techniques \cite{che2011dymaxion} and unified memory models \cite{landaverde2014investigation}, data transfer times between CPU cores and processing elements (PEs) have become a critical factor influencing overall execution time. While some prior work—particularly in the context of soft real-time scheduling—has considered data copy overheads, emerging real-time artificial intelligence workloads, such as those dominated by general matrix multiplication (GEMM), often experience substantial delays due to data transfers \cite{brightwell2018resource}.

Moreover, processor vendors are increasingly incorporating dedicated data copy engines to support multiple parallel transfers between CPUs and PEs \cite{choquette2021nvidia}, further underscoring the importance of modeling and optimizing memory transfers. Consequently, data copy across processor memories via the memory bus should no longer be treated as negligible. Future research should focus on accurately modeling data copy times within heterogeneous architectures and designing corresponding scheduling strategies that account for these overheads.

\subsection{Scheduling Policies with Response Time Analysis}
For soft real-time tasks, researchers have proposed numerous heuristic scheduling approaches, as precise response time analysis and schedulability guarantees are not strictly required.
In contrast, hard real-time tasks typically rely on classical schedulers such as Rate Monotonic (RM) and Earliest Deadline First (EDF). Many studies extend these classic scheduling policies by incorporating tailored designs either on CPUs or accelerators. These approaches are often grounded in the classical schedulers because they come with well-established response time analyses and schedulability tests. However, there is a notable lack of co-designed heuristic metrics that simultaneously consider both CPUs and accelerators, underscoring the difficulty of scheduling and analyzing response times in heterogeneous systems. Furthermore, to date, no universal scheduler has been identified that can effectively handle all types of heterogeneous architectures. This is primarily due to the wide variability in processor types, quantities, execution patterns, and optimization objectives across different systems. As a result, designing general-purpose real-time schedulers, especially with tight or even exact response time analysis and schedulability tests, for heterogeneous environments remains an open and challenging research problem.

\subsection{Fair and Standardized Evaluations}
Not only application-driven designs, but also general real-time scheduling strategies for soft or hard real-time tasks, face significant challenges in performing fair “apple-to-apple” comparisons. Currently, there is a lack of standardized metrics, methodologies, or frameworks to enable objective evaluation across different approaches. One reason for this is that accelerator-based heterogeneous architectures exhibit diverse execution patterns, as summarized in Sec. \ref{sec:model}. Another contributing factor is the absence of widely accepted metrics and evaluation methods for assessing the real-time performance of such architectures. 
For example, even for the response time analysis works which based on the same model, the evaluation metrics can be different. For example, studies such as \cite{gunzel2022edf,huang2016self} define the utilization rate as the total workload on CPU cores divided by the number of CPU cores:
$U = \frac{\sum\sum C_{i}^{j}}{N_{\text{CPU}}},$
based on the assumption that CPU cores are more dominant in the system. These works typically consider CPU cores as the host and treat accelerators as subordinate devices.
In contrast, other studies such as \cite{xu2022shape,zou2021rtgpu} define utilization by combining the workloads of both CPUs and accelerators, normalized by the total number of processing elements:
$
U = \frac{\sum\sum \left(C_{i}^{j} + A_{i}^{j}\right)}{N_{\text{CPU}} + N_{\text{accelerator}}},
$
reflecting the perspective that CPU cores and accelerators are equally important components of the system.
As the advances of research on real-time scheduling of accelerator-based heterogeneous architectures, fair and standardized evaluations are necessary.

To summarize, with the growing popularity of artificial intelligence (AI)-based time-critical applications and the widespread adoption of accelerator-based heterogeneous architectures as mainstream computing platforms over the past decade, real-time scheduling on such architectures has become increasingly important and has attracted significant attention. This survey provides a comprehensive overview of architectural features, execution models, and corresponding scheduling approaches, and concludes with key challenges and open research questions. We hope this survey serves both as an accessible introduction for beginners and a valuable reference for researchers and industry practitioners.

\bibliographystyle{unsrt}
\bibliography{bibliography}

\begin{IEEEbiography}[{\includegraphics[width=1in,height=1.25in,clip,keepaspectratio]{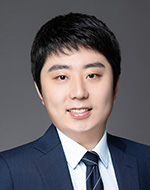}}]{An Zou} (Senior Member, IEEE) is an Associate Professor at the University of Michigan-Shanghai Jiao Tong University Joint Institute. His research focuses on computer architecture, embedded systems, processor low-power design. Dr. An Zou received his Ph.D. degree in Electrical Engineering from Washington University in St. Louis in 2021 and his M.S. and B.S. degrees from Harbin Institute of Technology in 2015 and 2013. He led or participated in several research projects and industry projects. His work has been extensively published and recognized at top-tier conferences and journals. He serves as the TPC of RTSS, DAC, ICCAD, and DATE. He was a recipient of Shang A. Richard Newton Young Student Fellow Award, Best Paper Nominations at DAC 2017, MLCAD 2020. 
\end{IEEEbiography}

\begin{IEEEbiography}[{\includegraphics[width=1in,height=1.25in,clip,keepaspectratio]{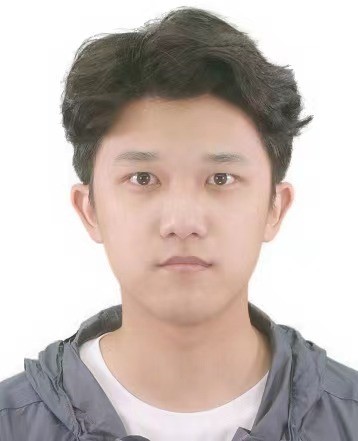}}]{Yuankai Xu} is a Ph.D. student at the Electrical and Computer Engineering Department in the University of Michigan - Shanghai Jiao Tong University Joint Institute. He received his B.S. degree from Shanghai Jiao Tong University in 2023. He is currently working on computer architectures, like low-power computing, real-time scheduling on heterogeneous computing architectures. He received the first prize award in the ACM SIGBED student research competition in 2022.
\end{IEEEbiography}
\vspace{-1cm}
\begin{IEEEbiography}[{\includegraphics[width=1in,height=1.25in,clip,keepaspectratio]{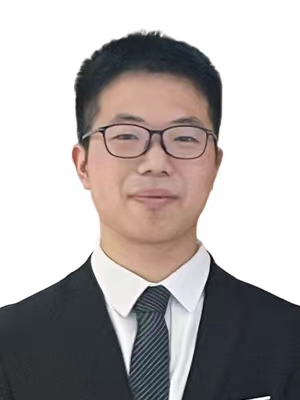}}]{Yinchen Ni} is a graduate student majoring in  Computer Science and Technology at the University of Michigan - Shanghai Jiao Tong University Joint Institute. He is expected to receive his B.S. degree from Shanghai Jiao Tong University in 2024. He is interested in the research real-time scheduling on heterogeneous computing architectures and timing-critical computing systems.
\end{IEEEbiography}
\vspace{-1cm}
\begin{IEEEbiography}[{\includegraphics[width=1in,height=1.25in,clip,keepaspectratio]{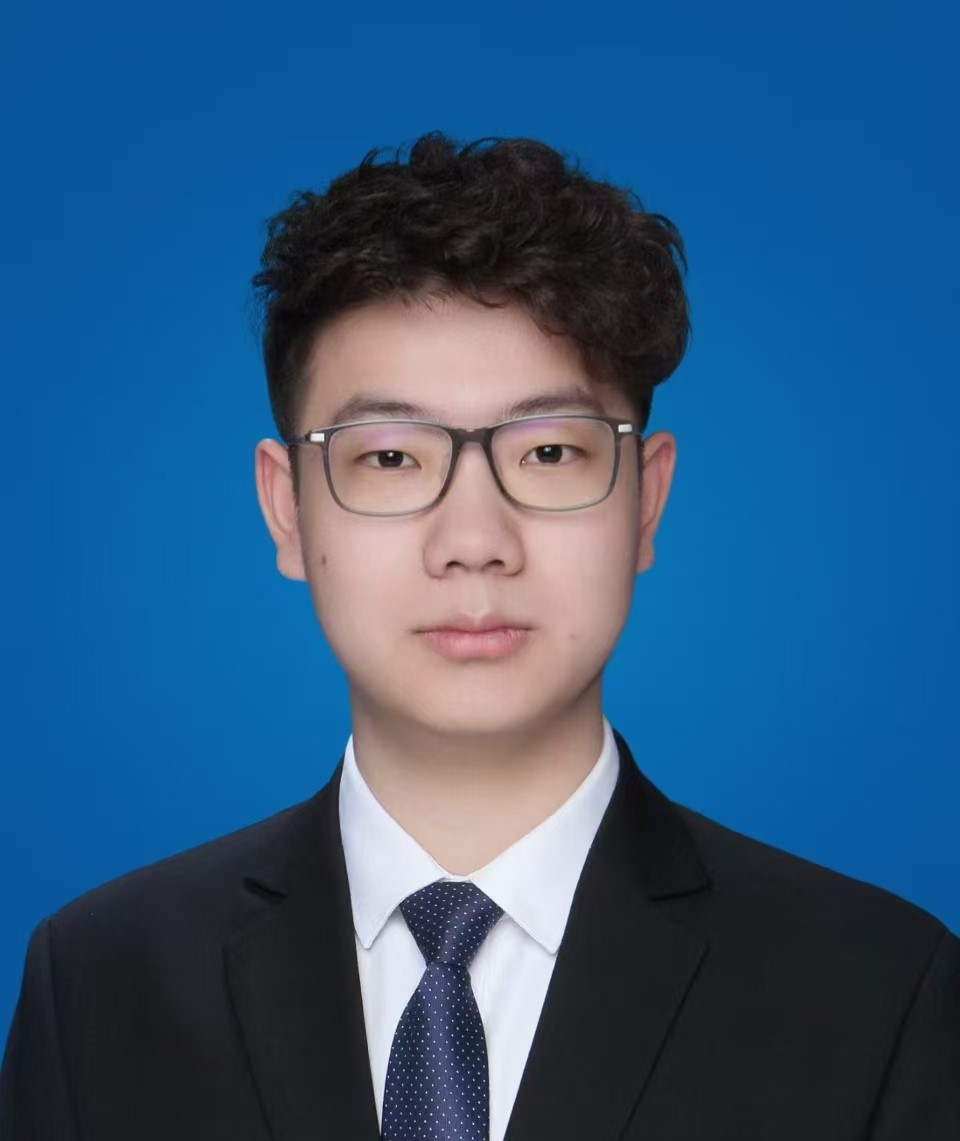}}]{Jintao Chen} Jintao Chen received the B.S. degree in automatic control (IEEE class) from Shanghai Jiao Tong University, Shanghai, China, in 2023. He is currently pursuing a Ph.D. degree in the Department of Automation, School of Electronic Information and Electrical Engineering, Shanghai Jiao Tong University, Shanghai, China. His areas of interest include real-time scheduling, real-time control, and intelligent control.
\end{IEEEbiography}
\vspace{-1cm}
\begin{IEEEbiography}[{\includegraphics[width=1in,height=1.25in,clip,keepaspectratio]{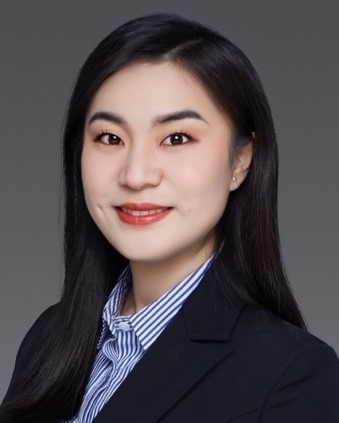}}]{Yehan Ma} (Senior Member, IEEE) is an Associate Professor in the School of Computer Science at Shanghai Jiao Tong University. Her research focuses on the cyber-physical systems, real-time control systems, and edge computing. She broadly investigated techniques and solutions for holistic managements of computation, communication, and control in cyber-physical systems. Dr. Ma received her Ph.D. degree in Computer Science from Washington University in St. Louis in 2020 and her M.S. and B.S. degrees from Harbin Institute of Technology in 2015 and 2013. Her work has been published at top-tier conferences and journals, such as RTSS, RTAS, EMSOFT, ICCPS, and TCAD. She received the TASE 2024 Best Application Paper Award. She serves as the TPC of RTSS, RTAS, ECRTS, and SECON.
\vspace{-1mm}
\end{IEEEbiography}

\vspace{-1cm}
\begin{IEEEbiography}[{\includegraphics[width=1in,height=1.3in,clip,keepaspectratio]{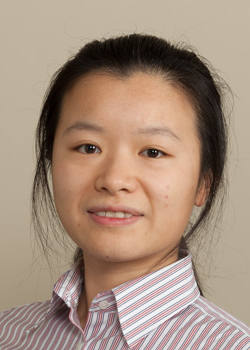}}]{Jing Li} 
is an Associate Professor in the Department of Computer Science at New Jersey Institute of Technology. She received her Ph.D. degree from Washington University in St. Louis in 2017. Her research interests include real-time systems, parallel computing, and reinforcement learning for system design and optimization. She has high impact publications in top conferences with three outstanding paper awards. Jing is the recipient of the NSF CAREER Award in 2024 and  Department of Energy Early Career Research Program Award in 2023.
\end{IEEEbiography}

\vspace{-1cm}
\begin{IEEEbiography}[{\includegraphics[width=1in,height=1.25in,clip,keepaspectratio]{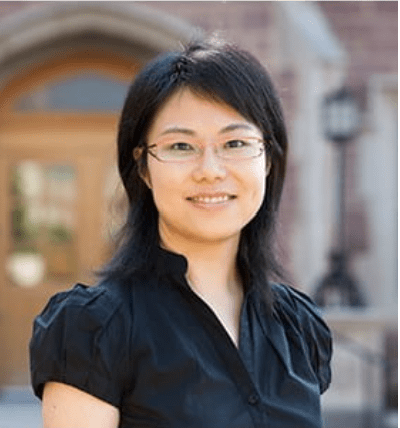}}]{Xuan Zhang} is an Associate Professor in the Electrical and Computer Engineering Department at Northeastern University. She works across the fields of VLSI design, computer architecture, and cyberphysical systems and her research interests include hardware/software co-design for efficient machine learning and artificial intelligence, real-time computing for autonomous systems in analog/mixed-signal and physical domain. Before joining Washington University, Dr. Zhang was a Postdoctoral Fellow in Computer Science at Harvard University. She received her BE degree in Electrical Engineering from Tsinghua University in China, and her MS and Ph.D. degrees in Electrical and Computer Engineering from Cornell University. Dr. Zhang is the recipient of NSF CAREER Award in 2020, AsianHOST Best Paper Award in 2020, DATE Best Paper Award in 2019, and ISLPED Design Contest Award in 2013, and her work has also been nominated for Best Paper Awards at ASP-DAC 2021, DATE 2019 and DAC 2017.
\vspace{-1mm}
\end{IEEEbiography}

\vspace{-1cm}
\begin{IEEEbiography}[{\includegraphics[width=1in,height=1.25in,clip,keepaspectratio]{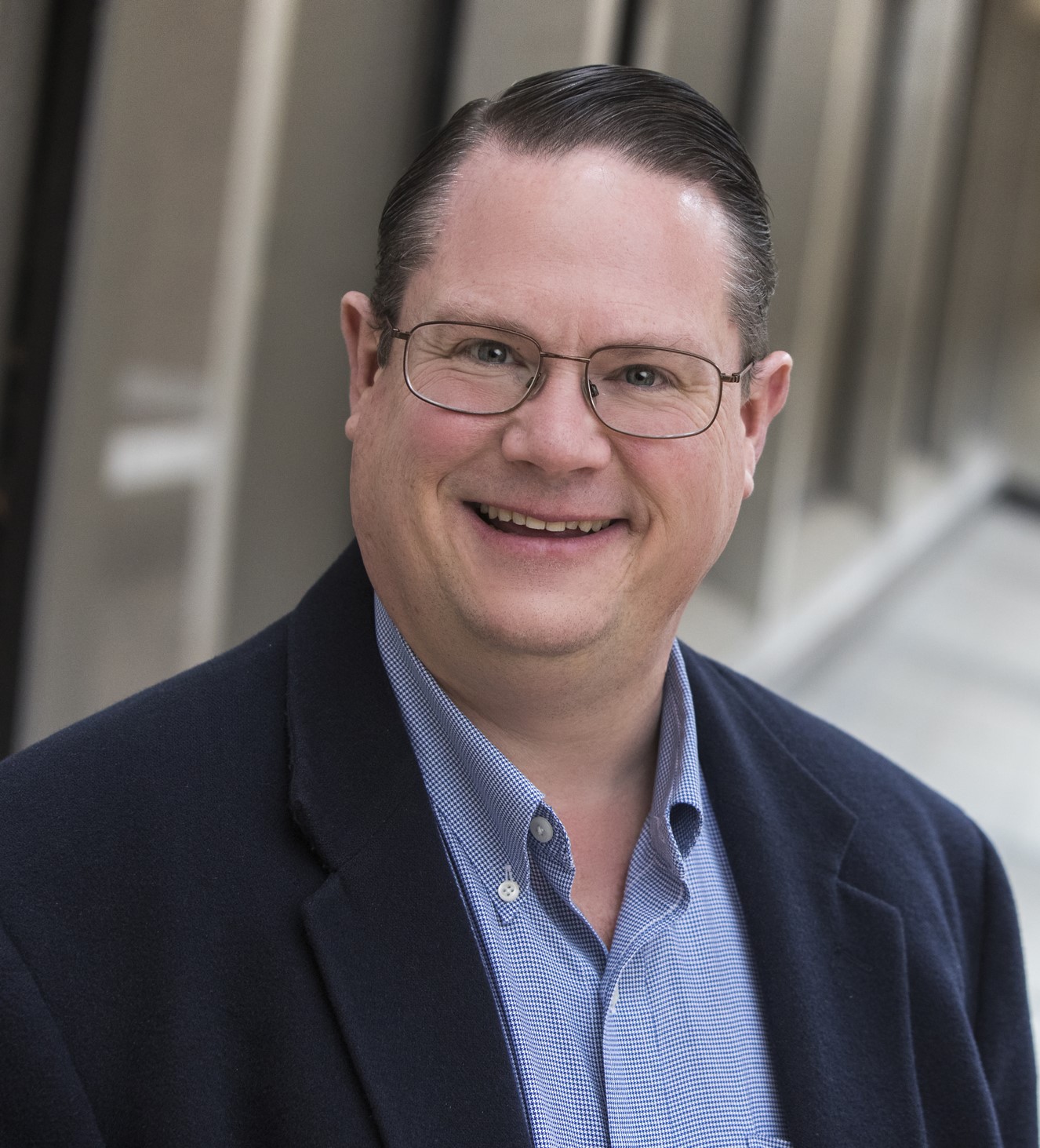}}]{Christopher Gill} is a Professor in the Department of Computer Science and Engineering at Washington University in St. Louis. He has published more than 100 technical articles in selective peer-reviewed conferences and journals, and has led or contributed to the development, evaluation, and open-source release of numerous real-time systems research platforms and artifacts, including: the Kokyu real-time scheduling and dispatching framework that was used in several AFRL and DARPA projects and flight demonstrations; the nORB small-footprint real-time object request broker; a number of real-time and fault-tolerant services for The ACE ORB (TAO) and the Component Integrated ACE ORB (CIAO); the Cyberphysical Instrument for Real-time hybrid Structural Testing (CIRST) that established key foundations for real-time hybrid simulation (RTHS), and the CyberMech platform that built on the CIRST project to enable parallel RTHS at millisecond time scales; and the RT-Xen real-time virtualization research platform and the RTDS scheduler that is now part of the Xen open-source software distribution. Professor Gill has served as an Associate Editor for TCPS and Subject Area Editor for the Elsevier Journal of Systems Architecture. He has served in numerous other organizing and technical reviewing roles within the real-time systems research community, including: IEEE TCRTS Chair; IEEE TCRTS ViceChair; IEEE RTSS General Chair; ACM SIGBED Vice-Chair; IEEE RTSS Technical Program Committee Chair; IEEE TCRTS Treasurer and IEEE RTSS Finance Chair.
\vspace{-1mm}
\end{IEEEbiography}

\vspace{-1cm}
\begin{IEEEbiography}[{\includegraphics[width=1in,height=1.25in,clip,keepaspectratio]{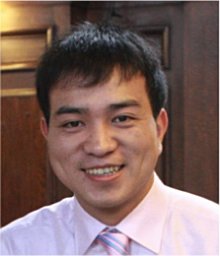}}]{Yier Jin} (Senior Member, IEEE) received the Ph.D. degree in electrical engineering from Yale University in 2012. He is currently a professor at the University of Science and Technology of China and an adjacent professor at the University of Florida. His research interests include hardware security, embedded systems design and security, trusted hardware intellectual property (IP) cores, hardware software co-design for modern computing systems, security analysis on the Internet of Things (IoT), and wearable devices with particular emphasis on information integrity and privacy protection in the IoT era. He was a recipient of the DoE Early CAREER Award in 2016 and the ONR Young Investigator Award in 2019. He received Best Paper Award at DAC’15, ASP-DAC’16, HOST’17, ACM TODAES’18, GLSVLSI’18, and DATE’19. 
\vspace{-1mm}
\end{IEEEbiography}

\end{document}